\documentclass{article}

\usepackage[main,final]{neurips_2026}

\usepackage[utf8]{inputenc}
\usepackage[T1]{fontenc}
\usepackage{hyperref}
\hypersetup{hidelinks}
\usepackage{url}
\usepackage{booktabs}
\usepackage{multirow}
\usepackage{caption}
\usepackage{subcaption}
\usepackage{listings}
\usepackage{amsfonts}
\usepackage{amssymb}
\usepackage{enumitem}
\usepackage{amsmath}
\usepackage{bm}
\usepackage{nicefrac}
\usepackage{microtype}
\usepackage{graphicx}
\usepackage[table]{xcolor}

\usepackage[most]{tcolorbox}
\usepackage{listings}
\usepackage{xcolor}

\definecolor{promptblue}{RGB}{34,102,170}
\definecolor{promptbg}{RGB}{245,248,252}
\definecolor{promptframe}{RGB}{180,195,215}

\lstdefinestyle{promptstyle}{
  basicstyle=\ttfamily\footnotesize,
  breaklines=true,
  columns=fullflexible,
  keepspaces=true,
  showstringspaces=false,
  frame=none
}

\newtcolorbox{promptbox}[1]{
  enhanced,
  breakable,
  colback=promptbg,
  colframe=promptframe,
  boxrule=0.8pt,
  arc=2pt,
  left=6pt,
  right=6pt,
  top=6pt,
  bottom=6pt,
  title=#1,
  coltitle=black,
  fonttitle=\bfseries,
  colbacktitle=promptbg,
  attach boxed title to top left={xshift=2mm,yshift*=-2mm},
  boxed title style={
    colback=white,
    colframe=promptblue,
    boxrule=0.8pt,
    arc=2pt
  }
}

\title{Can Complementary Signals Bridge Similarity Islands? Manifold-Augmented Graph Embedding for Multimodal Recommendation}

\author{
\textbf{Zhenye Yang}$^{1,2}$ \enspace \textbf{Jinpeng Chen}$^{1,2}$\thanks{Corresponding author} \enspace \textbf{Wenbo Fu}$^{1,2}$ \enspace \textbf{Jun Ma}$^{1,2}$ \enspace \textbf{Huan Li}$^{3}$ \\[3pt]
\textbf{Qimin Zhou}$^{1,2}$ \quad \textbf{Tao Wang}$^{1,2}$ \quad \textbf{Li Sun}$^{1}$ \quad \textbf{Senzhang Wang}$^{4}$ \\[3pt]
  {\normalfont\small $^{1}$School of Computer Science (National Pilot Software Engineering School),} \\
  {\normalfont\small Beijing University of Posts and Telecommunications, Beijing, China} \\
  {\normalfont\small $^{2}$Key Laboratory of Trustworthy Distributed Computing and Service (BUPT), Ministry of Education} \\
  {\normalfont\small $^{3}$Zhejiang University, Hangzhou, China \quad $^{4}$Central South University, Changsha, China} \\
  {\normalfont\small \{yangzhenye,jpchen,clearmylove,majun,zhouqimin,wangtao005,lsun\}@bupt.edu.cn} \\
  {\normalfont\small lihuan.cs@zju.edu.cn, szwang@csu.edu.cn}
}

\begin{document}

\maketitle

\begin{abstract}
Multimodal Recommender Systems (MMRS) leverage rich item features to mitigate data sparsity, yet similarity-driven graph construction and propagation can favor semantically similar substitutes over functionally related complements. When behavioral associations connect semantically distant items, complementary signals can remain geometrically isolated, yielding fragmented ``Similarity Islands'' and semantic drift when standard propagation is applied. To overcome these limitations, we propose \textbf{MAGE} (\underline{M}anifold-\underline{A}ugmented \underline{G}raph \underline{E}mbedding), a framework for personalized multimodal ranking that decouples complementarity modeling into offline geometric warping and online topological propagation. 
The offline phase enhances item representations by extracting complement-aware residual directions via a Soft Conceptor mechanism and injecting them under a local gap-preserving constraint, incorporating co-purchase evidence while controlling the magnitude of geometric perturbation. Synergistically, the online phase employs an asymmetric dual-stream GNN that treats visual features as stable anchors and augmented textual features as flexible probes, fused through a self-calibrated cross-modal mechanism to reconcile shared semantics and modality-specific residual cues. Extensive experiments on four Amazon datasets show that MAGE yields consistent improvements over strong multimodal recommendation baselines. Together with the supporting analyses in the appendix, these results suggest that geometry-aware augmentation can alleviate the semantic mismatch that arises in complementary recommendation settings. Our code is available at \url{https://github.com/Alex-bupt/mage}.

\end{abstract}

\section{Introduction}

Recommender systems help users navigate large collections of online information~\cite{zhang2019deep}. 
While traditional collaborative filtering struggles with persistent \textit{data sparsity}~\cite{natarajan2020resolving}, Multimodal Recommender Systems (MMRS) have emerged as a powerful remedy. 
By integrating rich side information such as product images and textual descriptions, MMRS construct dense semantic representations for items, substantially improving recommendation quality, particularly in cold-start and long-tail scenarios~\cite{liu2024multimodal, chen2017attentive, lin2024gume}.

Graph Neural Networks (GNNs) have become the dominant paradigm in MMRS due to their ability to propagate user preferences through higher-order connectivity~\cite{gao2023survey}. 
Models such as MMGCN~\cite{wei2019mmgcn}, MENTOR~\cite{xu2025mentor}, and GUME~\cite{lin2024gume} refine graph structures and multimodal fusion. 
However, similarity-driven graph construction and propagation favor the \textbf{homophily assumption}~\cite{zhu2020beyond}. 
Low-pass propagation~\cite{chienadaptive} encourages representation smoothness across connected nodes, implicitly assuming that ``connected items should be similar''~\cite{platonov2023characterizing}. 
This favors \textit{substitutes}, such as two similar phone models, that naturally cluster in the embedding space.

Nevertheless, real-world purchasing behavior is frequently driven by \textbf{Complementarity} (e.g., co-purchase patterns), where users seek items that are functionally compatible rather than semantically similar (e.g., a \textit{Charger Plug} for an iPhone)~\cite{mcauley2015inferring}. 
Although complementary recommendation has been studied using behavioral signals and graph-based formulations~\cite{shang2024transitivity, luo2024spectral}, similarity-driven multimodal approaches can retain a mismatch between behavioral topology and item features even after adding behavioral edges. Such designs become particularly limited in \textbf{multimodal settings}, where complementary items often differ drastically in both visual appearance and textual descriptions, giving rise to pronounced \textbf{heterophily}~\cite{zheng2022graph}. In contrast to generic heterophily-aware graph learning, our setting must additionally reconcile modality conflict: co-purchase relations provide imperfect evidence of functional relatedness, while the raw feature space remains organized around semantic similarity. This mismatch yields fragmented ``Similarity Islands'': substitute items cluster naturally, while behaviorally associated complements remain geometrically separated and difficult to connect through similarity-driven propagation. Empirical statistics and visualizations supporting the existence of these ``Similarity Islands'' are provided in Appendix~\ref{app:similarity_islands}. Uniform smoothing over such heterophilic relations can then introduce \textbf{semantic drift}~\cite{rusch2023survey}, because enforcing smoothness between semantically distant but behaviorally related items may contaminate the seed item's core representation and weaken complementary exploration.

To better address these ``Similarity Islands'' without exacerbating semantic drift, we argue that multimodal recommendation with complementary signals may benefit from \emph{geometric transformation} in addition to topological augmentation. We therefore propose \textbf{MAGE} (\textbf{M}anifold-\textbf{A}ugmented \textbf{G}raph \textbf{E}mbedding), a framework that decouples complementary modeling into two tightly coupled components: offline geometric augmentation and online asymmetric propagation. Our task is personalized user--item ranking, rather than recommending complements for a query item alone. At the representation level, MAGE performs offline manifold-based enhancement by extracting complement-aware residual directions via local SVD and Soft Conceptor projection, and injecting them into the textual modality under an explicit local gap-preserving constraint. At the architectural level, MAGE adopts an \textbf{asymmetric dual-stream} design~\cite{wang2020disentangled}, treating visual features as stable perceptual anchors and augmented textual features as flexible functional probes. A \textbf{Self-Calibrated Gate} is further introduced to harmonize these heterogeneous signals, so that complementary information can be incorporated without overwhelming an item's core identity.

\begin{enumerate}[label=\roman*,
                  leftmargin=2.2em,
                  labelsep=0.6em,
                  itemsep=0.6em,
                  topsep=0.2em]
   \item We study the representation--topology mismatch of complementary signals in personalized multimodal recommendation. Building on existing complementarity and heterophily research, we introduce \textbf{geometry-aware manifold reshaping} before topological propagation.
   \item We propose MAGE, a framework that combines offline geometric augmentation with online asymmetric propagation. In Phase~I, complement-aware residuals are injected under an explicit local gap-preserving rule with an adaptive coefficient; in Phase~II, the \textit{Self-Calibrated Gate} harmonizes stable visual anchors with augmented textual signals.
   \item Experiments on four benchmark datasets show consistent improvements over strong baselines. A targeted hard-complement evaluation, fine-grained ablations, original-text controls, and graph perturbations characterize the contributions of geometric augmentation, modality-specific propagation, and text preprocessing, as well as sensitivity to imperfect graph evidence.
\end{enumerate}

\section{Related Work}
\paragraph{Multimodal graph recommendation.}
VBPR integrates visual features with collaborative ranking~\cite{he2016vbpr}; MMGCN, MGCN, LGMRec, MENTOR, and GUME develop multimodal propagation, representation learning, and fusion~\cite{wei2019mmgcn,yu2023multi,guo2024lgmrec,xu2025mentor,lin2024gume}. MAGE builds on this setting but explicitly transforms the textual geometry before assigning similarity and complementary connections to different streams. 
\paragraph{Complementarity and heterophily.}
Prior complementary-item recommendation already models relevance together with dissimilarity~\cite{mcauley2015inferring,luo2024spectral,shang2024transitivity}. In particular, SComGNN uses low- and mid-pass spectral filters to capture these two attributes~\cite{luo2024spectral}. MAGE instead targets personalized user--item ranking with multimodal inputs, coupling a local feature transformation to modality-specific propagation. Generic heterophily-aware learning motivates departing from uniform smoothing~\cite{zhu2020beyond,zheng2022graph,pei2020geom}; here the additional issue is how behavioral topology interacts differently with visual and textual features.

\section{Problem Definition}

Let $\mathcal{U}$ and $\mathcal{I}$ denote the sets of users and items. Each item $i\in\mathcal{I}$ is associated with multimodal features: a $d_v$-dimensional visual embedding $\mathbf{x}_{i,v}\in\mathbb{R}^{d_v}$ and a $d_t$-dimensional textual embedding $\mathbf{x}_{i,t}\in\mathbb{R}^{d_t}$. User--item interactions are given by a binary matrix $\mathbf{R}\in\{0,1\}^{|\mathcal{U}|\times|\mathcal{I}|}$ with $R_{u,i}=1$ indicating an observed interaction. We further leverage co-purchase relations $\mathcal{G}_{\mathrm{co}}=(\mathcal{I},\mathcal{E}_{\mathrm{co}})$, where $(a,b)\in\mathcal{E}_{\mathrm{co}}$ denotes an observed co-purchase association, used as imperfect evidence of complementarity. The graph comes from item metadata, not held-out user--item labels.

Similarity-based propagation encourages connected items to have similar representations. However, edges in $\mathcal{E}_{\mathrm{co}}$ are often heterophilic: complementary items are functionally related but semantically distant in the raw feature space, making naive propagation prone to semantic drift.

Our goal is personalized user--item ranking with score $\hat{y}_{u,i}$. Specifically, we seek a \textbf{manifold transformation} $\phi(\cdot)$ that maps $\mathbf{x}_{i,t}$ to an augmented space $\tilde{\mathbf{x}}_{i,t}$, leveraging locality-preserving geometry to capture complementary signals beyond the dominant semantic subspace~\cite{he2003locality, pei2020geom}. The final model combines $\tilde{\mathbf{x}}_{i,t}$ with the stable visual anchors $\mathbf{x}_{i,v}$ to model both substitute and complementary preferences.

\section{Methodology}
\begin{figure*}[ht]
    \centering
    \includegraphics[width=\textwidth]{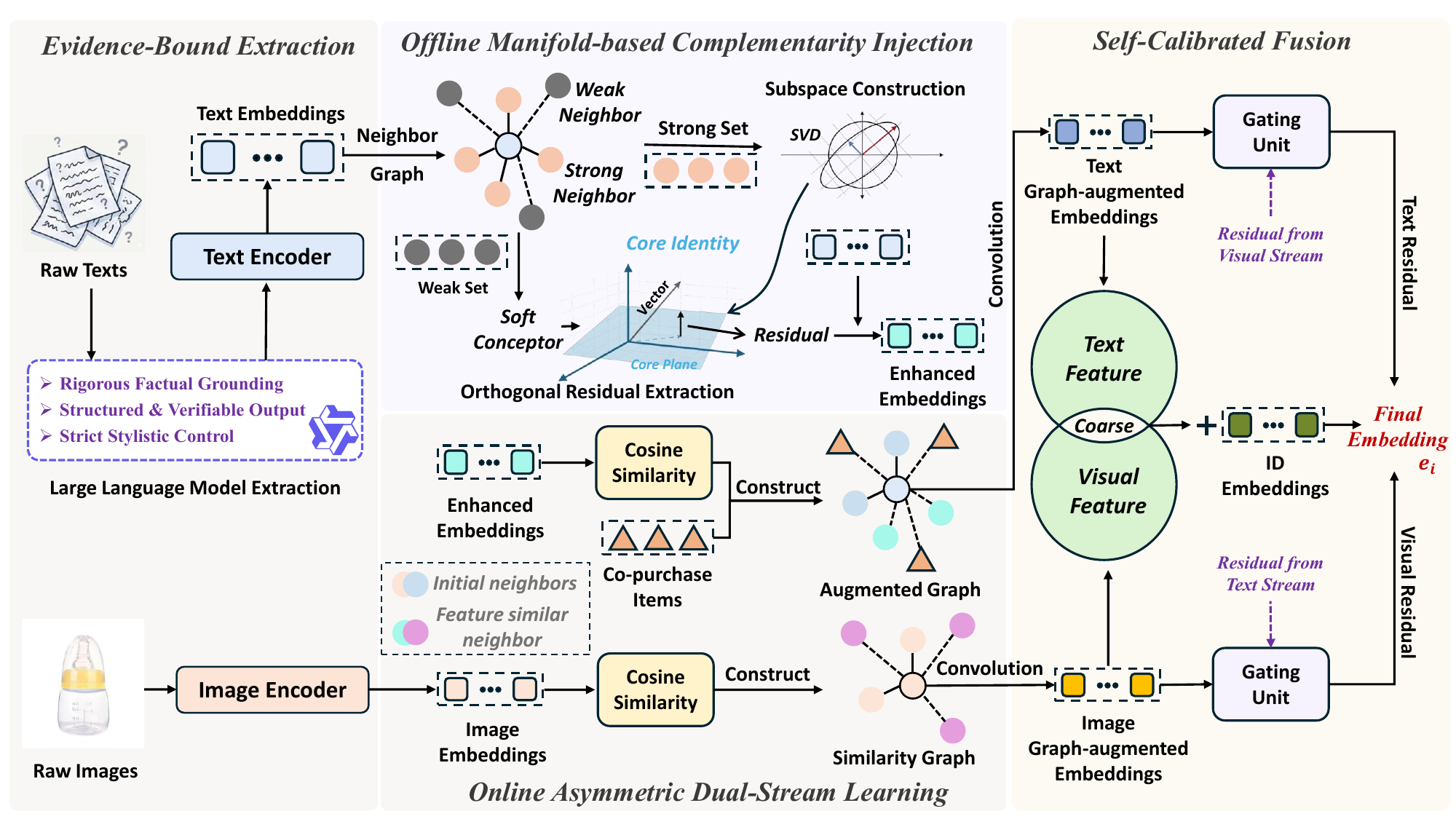}
    \caption{Architecture of the MAGE framework.}
    \label{fig:framework}
\end{figure*}

In this section, we present MAGE, a framework designed to bridge the semantic gap in multimodal recommendation by leveraging complementary signals.
The core philosophy of MAGE is to decouple the challenge of modeling \textbf{heterophilic complementary signals} into two distinct spaces:
(1) the \textbf{Geometric Space}, where we resolve feature conflicts offline via manifold learning; and
(2) the \textbf{Topological Space}, where we propagate complementary signals online via an asymmetric graph neural network.
An overview of the proposed MAGE framework is illustrated in Figure~\ref{fig:framework}.

\subsection{Modality Initialization: Evidence-Bound Extraction}
\label{sec:initialization}

Raw product text in e-commerce is often noisy and inconsistently formatted, making functional cues such as compatibility constraints and specifications difficult to identify directly. We therefore apply a one-time \textbf{LLM-based evidence-bound extraction} step as data preprocessing~\cite{wei2024llmrec}, which is instructed to extract explicitly supported functional attributes and remove marketing-heavy content. The processed text is then encoded into a $d$-dimensional embedding $\mathbf{x}_{i,t}$, which serves as the textual input for subsequent manifold augmentation. The extracted descriptions are frozen before representation learning; no LLM is called during training or inference. Detailed prompts and failure handling are provided in Appendices~\ref{app:prompt} and~\ref{app:preprocessing_protocol}, respectively, and the original-text control is reported in Section~\ref{sec:attribution}.

\subsection{Phase~I: Offline Manifold-based Complementarity Injection}
\label{sec:phase1}

Complementary items are often geometrically distant in the raw feature space, making direct aggregation prone to noise and semantic drift. Phase~I therefore performs an offline geometric transformation on the textual modality, mapping $\mathbf{x}_{i,t}$ to $\tilde{\mathbf{x}}_{i,t}$ before graph propagation. The key idea is to extract a \emph{complement-aware residual direction} from the local co-purchase neighborhood, while constraining its injection so that the original substitute--complement margin is not overly distorted.

\subsubsection{Local Relative Subspace Characterization}

Given a seed item $i$, we consider its co-purchase neighborhood $\mathcal{N}_{full}(i)$ and compute cosine similarities between the seed embedding $\mathbf{x}_{i,t}$ and its neighbors. Instead of using a global threshold shared across all items, we adopt a \textbf{per-seed relative percentile partition}: neighbors in the upper tail of the local similarity distribution form the \textit{Strong Set} $\mathcal{S}_i$, while those in the lower tail form the \textit{Weak Set} $\mathcal{W}_i$. These sets are local geometric proxies for substitute-like and complement-like relations, not verified semantic labels. To further suppress noisy co-purchase relations, we apply an additional similarity floor to weak neighbors, excluding extremely dissimilar items that are unlikely to encode meaningful complementarity.

We then build a local semantic matrix from the seed item and its strong neighbors,
\begin{equation}
    \mathbf{C}_i = \begin{bmatrix}
    \mathbf{x}_{i,t} \\
    \{\mathbf{x}_{j,t}\}_{j\in\mathcal{S}_i}
    \end{bmatrix},
    \qquad
    \mathbf{C}_i = \mathbf{U}_i \boldsymbol{\Sigma}_i \mathbf{V}_i^\top,
\end{equation}
where $\mathbf{V}_i$ spans the dominant substitute subspace around item $i$. Rather than performing a hard truncation on singular directions, we use a \textbf{Soft Conceptor} to softly characterize this subspace:
\begin{equation}
    w_{i,k} = \frac{\sigma_k^2}{\sigma_k^2 + \gamma^{-2}},
\end{equation}
where $\sigma_k$ is the $k$-th singular value and $\gamma$ controls the softness of the subspace boundary. Stronger semantic directions receive larger weights and are thus more aggressively suppressed in subsequent residual extraction.

\subsubsection{Complement Residual Extraction}

We use deviations from the dominant substitute-like subspace as candidate complementary signals. Co-purchase evidence and geometric constraints motivate this proxy, but do not certify that each residual is semantically complementary. Let $\mathbf{V}_i=[\mathbf{v}_{i,1},\ldots,\mathbf{v}_{i,r}]$ denote the right singular vectors and $\mathbf{w}_i=[w_{i,1},\ldots,w_{i,r}]$ the corresponding Soft Conceptor weights. We define the soft projection operator as
\begin{equation}
    \mathbf{P}_i
    =
    \mathbf{V}_i \mathrm{diag}(\mathbf{w}_i)\mathbf{V}_i^\top,
\end{equation}
and remove the substitute-dominated component from each weak neighbor. Aggregating these filtered signals yields a consensus complementary direction:
\begin{equation}
    \bar{\mathbf{r}}_i
    =
    \frac{1}{|\mathcal{W}_i|}
    \sum_{j\in\mathcal{W}_i}
    (\mathbf{I}-\mathbf{P}_i)\mathbf{x}_{j,t}.
\end{equation}
To remove the component aligned with the seed representation, we orthogonalize this consensus direction against the unit-normalized seed embedding:
\begin{equation}
    \mathbf{r}_i
    =
    \bar{\mathbf{r}}_i
    -
    (\bar{\mathbf{r}}_i^\top \mathbf{x}_{i,t})\,\mathbf{x}_{i,t}.
\end{equation}
The resulting $\mathbf{r}_i$ serves as the final complement-aware residual direction of item $i$.

\subsubsection{Gap-Constrained Adaptive Injection}
A major risk of residual enhancement is over-injection: a large update may collapse the local separation between substitute-like and complement-like neighbors. We therefore adapt the injection strength to each seed rather than using a fixed global coefficient. The constraint concerns the strong/weak geometric proxies defined above; it does not assign semantic labels to individual neighbors.

Let $\boldsymbol{\mu}_{\mathcal S_i}$ and $\boldsymbol{\mu}_{\mathcal W_i}$ be the strong- and weak-set means, and define $\boldsymbol{\Delta}_i=\boldsymbol{\mu}_{\mathcal S_i}-\boldsymbol{\mu}_{\mathcal W_i}$, $g_i=\mathbf{x}_{i,t}^{\top}\boldsymbol{\Delta}_i$, and $d_i=\mathbf r_i^{\top}\boldsymbol{\Delta}_i$. Here the two means average the original textual embeddings in their respective sets. The value $g_i$ measures how much more closely the seed aligns with its strong neighbors, while $d_i$ measures how the residual changes this separation. For $g_i>0$, we set the target gap to $g_i^{\mathrm{target}}=\rho g_i$, where $\rho\in(0,1)$ is the keep ratio. For an unnormalized update $\mathbf z_i(\alpha)=\mathbf x_{i,t}+\alpha\mathbf r_i$, the requirement is
\begin{equation}
\mathbf z_i(\alpha)^\top\boldsymbol\Delta_i\geq\rho g_i.
\end{equation}
Since this update has gap $g_i+\alpha d_i$, for $g_i>0$ the initial coefficient is
\begin{equation}
\alpha_i=\begin{cases}
\min\!\left(\alpha_{\max},\dfrac{(1-\rho)g_i}{-d_i+\varepsilon}\right),&d_i<0,\\
\alpha_{\max},&d_i\geq0.
\end{cases}
\end{equation}
For $d_i<0$, a larger $\rho$ tightens the injection bound, while an active safety cap can leave the coefficient unchanged. For $d_i\geq0$, the unnormalized gap does not decrease, but the cap still limits perturbation. We then normalize $\tilde{\mathbf x}_{i,t}=(\mathbf x_{i,t}+\alpha_i\mathbf r_i)/\|\mathbf x_{i,t}+\alpha_i\mathbf r_i\|_2$ and backtrack $\alpha_i$ if $\tilde{\mathbf x}_{i,t}^{\top}\boldsymbol{\Delta}_i<\rho g_i$. The cap and normalized-gap check control injection magnitude; they do not prove semantic correctness. Appendix~\ref{app:gap_derivation} gives the expanded derivation, and Section~\ref{sec:attribution} isolates the safeguards empirically.

\subsection{Phase~II: Online Asymmetric Dual-Stream Learning}
\label{sec:phase2}

Because Phase~I changes textual features while leaving visual features intact, we use an \textbf{Asymmetric Dual-Stream} architecture for complement-aware multimodal recommendation, treating visuals as stable anchors and augmented text as flexible probes, and assigning stability to the visual stream and plasticity to the textual stream.

\subsubsection{Asymmetric Encoding Strategy}
We differentiate the roles of visual and textual modalities into ``Stability'' and ``Plasticity'' to balance identity preservation with complementary exploration:

\paragraph{Visual Anchor (Stability):} We treat the visual modality as a stable anchor to preserve the item's perceptual identity. We feed the \textit{raw} visual embeddings $\mathbf{x}_{i,v}$ into a non-linear space projector $\phi_v(\cdot)$, implemented as a multi-layer perceptron (MLP) with Sigmoid activation, to map them into the interaction latent space: $\mathbf{e}_{i,v}^{(0)} = \phi_v(\mathbf{x}_{i,v})$.

\paragraph{Textual Probe (Plasticity):} We treat the textual modality as a flexible probe to explore latent complementarity. Instead of raw text, we input the \textit{manifold-augmented} embeddings $\tilde{\mathbf{x}}_{i,t}$ into a separate encoder $\phi_t(\cdot)$ with distinct parameters: $\mathbf{e}_{i,t}^{(0)} = \phi_t(\tilde{\mathbf{x}}_{i,t})$. This effectively injects the offline-learned geometric residuals into the online learning pipeline.

\subsubsection{Modal-Asymmetric Topology Construction}
\label{sec:topology}

To decouple similarity from complementarity, we design two parallel graph convolution tracks with asymmetric connectivity. This design is motivated by our observation that visual features are prone to noise when propagating over complementary edges, whereas augmented textual features are assigned the role of carrying complementary information.

\paragraph{Visual Stream (Homophilic Topology):} 
This stream operates on a purified \textit{Similarity Graph} $\mathcal{G}_{\mathrm{sim}}$, constructed as the intersection of the visual $k$-NN graph and the textual $k$-NN graph. An item--item edge is retained only when it appears in both modality-specific nearest-neighbor graphs, so the resulting topology reflects visual--textual consensus without adding edges solely on the basis of co-purchase relations. The GCN on this track acts as a conservative smoother:
\begin{equation}
    \mathbf{H}_v^{(l+1)}
    =
    \sigma(\tilde{\mathbf{A}}_{\mathrm{sim}} \mathbf{H}_v^{(l)} \mathbf{W}_v^{(l)}),
\end{equation}
where $\tilde{\mathbf{A}}_{\mathrm{sim}}$ is the normalized adjacency matrix, $\mathbf{H}_v^{(l)}$ is the layer-$l$ visual item representation matrix, $\mathbf{W}_v^{(l)}$ is a learnable transformation, and $\sigma(\cdot)$ is a non-linear activation. This homophilic propagation keeps visual representations grounded in intrinsic category identity without being perturbed by heterophilic complementary edges.

\paragraph{Textual Stream (Hybrid Topology):} 
This stream operates on an \textit{Augmented Complementary Graph} $\mathcal{G}_{\mathrm{aug}}$. Here, we inject the filtered co-purchase edges $\mathcal{E}_{\mathrm{co}}$ into the $k$-NN structure. To maintain structural consistency and sparsity, we apply a unified truncation strategy where both similarity and complementarity are bounded by the same neighbor count $k$.

We define a weighted adjacency matrix $\mathbf{A}_{\mathrm{aug}}$ to balance the homophilic similarity signals and the heterophilic complementary residuals. For an edge $(i, j)$, the weight is assigned as follows:
\begin{equation}
    \mathbf{A}_{\mathrm{aug}}(i, j) = 
    \begin{cases} 
    1, & \text{if } (i, j) \in \mathcal{E}_{\mathrm{knn}} \\
    \omega_{\mathrm{co}}, & \text{if } (i, j) \in \mathcal{E}_{\mathrm{co}} \text{ and } \operatorname{rank}(j) \le k \\
    0, & \text{otherwise}
    \end{cases}
\end{equation}
Here, $\mathcal{E}_{\mathrm{knn}}$ denotes the edge set of the $k$-NN similarity graph, and $\operatorname{rank}(j)$ is the rank of neighbor $j$ among co-purchase candidates of item $i$ sorted by co-purchase strength. 
The hyperparameter $k$ bounds each edge source separately (their union can contain up to $2k$ neighbors before any symmetrization), while $\omega_{\mathrm{co}}$ serves as a signal modulator that down-weights complementary edges to prevent heterophilic propagation from overwhelming the item's core semantic identity. The textual GCN then performs a hybrid propagation:
\begin{equation}
    \mathbf{H}_t^{(l+1)}
    =
    \sigma(\tilde{\mathbf{A}}_{\mathrm{aug}} \mathbf{H}_t^{(l)} \mathbf{W}_t^{(l)}).
\end{equation}
By propagating the manifold-augmented textual features over this hybrid topology, MAGE diffuses high-order complementary signals while preserving the stability of the item representations.

In parallel, we maintain a collaborative ID backbone by propagating learnable user and item ID embeddings over the normalized user--item graph augmented with the textual item--item topology. The resulting representations, denoted as $\mathbf{e}_{u,\mathrm{id}}$ and $\mathbf{e}_{i,\mathrm{id}}$, serve as the behavioral backbone for fusion and contrastive regularization.

\subsection{Self-Calibrated Cross-Modal Fusion}
\label{sec:gate}
The streams contain shared category information and modality-specific cues. Our disentangle-then-fuse design estimates their common structure and calibrates the residuals before integration. From the final item streams $\mathbf h_{i,v}$ and $\mathbf h_{i,t}$, attention forms a shared prototype $\mathbf h_{i,c}=\alpha_v\mathbf h_{i,v}+\alpha_t\mathbf h_{i,t}$, where $(\alpha_v,\alpha_t)=\mathrm{Softmax}(\phi_{attn}(\mathbf h_{i,v})\parallel\phi_{attn}(\mathbf h_{i,t}))$. The collaborative ID representation modulates each residual: $\mathbf r_{i,m}=\sigma_{beh}^{(m)}(\mathbf e_{i,\mathrm{id}})\odot(\mathbf h_{i,m}-\mathbf h_{i,c})$. This separates deviations from the shared prototype and weights them using collaborative information; these residuals are not guaranteed to be purely complementary semantics. For $m\in\{v,t\}$ and its counterpart $\bar m$, the bidirectional gate refines it as
\begin{equation}
\mathbf g_{i,m}=\sigma\!\left(\mathrm{MLP}_{\bar m\to m}(\mathbf r_{i,\bar m})+\mathbf W_{m\to m}\mathbf r_{i,m}\right),\qquad
\tilde{\mathbf r}_{i,m}=\mathbf r_{i,m}+\beta_g\,\mathbf r_{i,m}\odot\mathbf g_{i,m}.
\end{equation}
The gate combines a cross-modal suggestion with self-modal calibration. Residual-style refinement preserves the input signal, with $\beta_g$ controlling the adjustment. The final item vector combines the ID backbone with the shared and refined signals:
\begin{equation}
\mathbf e_i=\mathbf e_{i,\mathrm{id}}+\mathbf e_{i,\mathrm{fused}},\qquad
\mathbf e_{i,\mathrm{fused}}=\tfrac13(\mathbf h_{i,c}+\tilde{\mathbf r}_{i,v}+\tilde{\mathbf r}_{i,t}).
\end{equation}
Users have no raw modality features. Their visual/textual views, $\mathbf e_{u,v}^{\mathrm{ext}}$ and $\mathbf e_{u,t}^{\mathrm{ext}}$, are induced by propagating item signals through interaction histories. Applying the same prototype/residual/gating rule to these history-induced views yields $\mathbf e_{u,\mathrm{fused}}$, giving $\mathbf e_u=\mathbf e_{u,\mathrm{id}}+\mathbf e_{u,\mathrm{fused}}$. Appendix~\ref{app:fusion_details} expands the computation.

\subsection{Prediction and Optimization}
\label{sec:optimization}

Given the final user and item representations $\mathbf{e}_u$ and $\mathbf{e}_i$, we predict the preference score by inner product:
\begin{equation}
    \hat{y}_{u,i} = \mathbf{e}_u^\top \mathbf{e}_i.
\end{equation}

We optimize MAGE with a multi-task objective
\begin{equation}
    \mathcal{L}
    =
    \mathcal{L}_{\mathrm{BPR}}
    +
    \lambda_{\mathrm{align}}\mathcal{L}_{\mathrm{align}}
    +
    \lambda_{\mathrm{bm}}\mathcal{L}_{\mathrm{BM}}
    +
    \lambda_{\mathrm{info}}\mathcal{L}_{\mathrm{UM}}
    +
    \lambda_{\mathrm{reg}}\|\Theta\|_2^2.
\end{equation}
Here, $\mathcal{L}_{\mathrm{BPR}}$ is the main recommendation objective, while the remaining terms act as regularizers. The Info Loss in the sensitivity analysis is $\mathcal{L}_{\mathrm{UM}}$ (\texttt{um\_loss}), weighted by $\lambda_{\mathrm{info}}$; $\mathcal{L}_{\mathrm{BM}}$ (\texttt{bm\_loss}) is a separate regularizer.

To reduce modality-wise scale mismatch, we match the moments of the stream embeddings $\mathbf E_v$ and $\mathbf E_t$. We also align the fused representations with the collaborative backbone:
\begin{align}
\mathcal L_{\mathrm{align}}&=|\mu(\mathbf E_v)-\mu(\mathbf E_t)|+|\sigma^2(\mathbf E_v)-\sigma^2(\mathbf E_t)|,\\
\mathcal L_{\mathrm{BM}}&=\mathcal L_{\mathrm{NCE}}(\mathbf e_{u,\mathrm{fused}},\mathbf e_{u,\mathrm{id}})+\mathcal L_{\mathrm{NCE}}(\mathbf e_{i,\mathrm{fused}},\mathbf e_{i,\mathrm{id}}).
\end{align}
For users, let $\mathbf e_{u,\mathrm{expl}}=\tfrac12(\mathbf e_{u,v}^{\mathrm{ext}}+\mathbf e_{u,t}^{\mathrm{ext}})$ denote the history-induced dual-stream view. The user regularizer combines agreement with this view and consistency between perturbed versions:
\begin{align}
\mathcal L_{\mathrm{UM}}={}&\mathcal L_{\mathrm{NCE}}(\mathbf e_{u,\mathrm{fused}},\mathbf e_{u,\mathrm{expl}})\\
&+\mathcal L_{\mathrm{NCE}}(\tilde{\mathbf e}_{u,\mathrm{fused}}^{(1)},\tilde{\mathbf e}_{u,\mathrm{fused}}^{(2)})+\mathcal L_{\mathrm{NCE}}(\tilde{\mathbf e}_{u,\mathrm{expl}}^{(1)},\tilde{\mathbf e}_{u,\mathrm{expl}}^{(2)}).\nonumber
\end{align}
These terms regularize representation learning alongside BPR; they are not independent complementarity objectives. Appendix~\ref{app:objectives} details the perturbation construction. 

\section{Experiments}

\subsection{Datasets and Experimental Inputs}
We conduct experiments on four public Amazon benchmark datasets: \textit{Baby}, \textit{Sports}, \textit{Clothing}, and \textit{Electronics}~\cite{yu2023multi, xu2025mentor, lin2024gume}. For the proposed pipeline, we preprocess raw product text with the \textbf{Evidence-Bound Extraction} pipeline using Qwen3-32B-AWQ~\cite{yang2025qwen3}, and encode the resulting descriptions into 512-dimensional textual features. Visual features are extracted as 4096-dimensional embeddings using a pre-trained encoder. The co-purchase graph is constructed from item-level metadata relations and kept separate from the user--item interaction labels used for evaluation.

To distinguish preprocessing effects from architectural gains, we evaluate baseline text replacement and MAGE with original text in Section~\ref{sec:attribution}. Dataset statistics, implementation settings, and sensitivity analyses are in Appendices~\ref{app:data_stats}, \ref{app:implementation}, and~\ref{app:hyperparams}.

\subsection{Compared Methods}
We compare MAGE with representative baselines from three categories: traditional collaborative filtering methods (\textbf{MF-BPR}~\cite{DBLP:conf/uai/RendleFGS09}, \textbf{LightGCN}~\cite{he2020lightgcn}), multimodal representation learning methods (\textbf{VBPR}~\cite{he2016vbpr}, \textbf{SLMRec}~\cite{tao2022self}, \textbf{BM3}~\cite{zhou2023bootstrap}, \textbf{MGCN}~\cite{yu2023multi}, \textbf{LGMRec}~\cite{guo2024lgmrec}), and recent graph structure learning methods that refine multimodal connectivity or cross-modal relations (\textbf{MENTOR}~\cite{xu2025mentor}, \textbf{GUME}~\cite{lin2024gume}).

\subsection{Overall Performance}
\label{sec:overall_performance}
Table~\ref{tab:main-results} reports full-test Recall and NDCG on four Amazon datasets. MAGE achieves the best reported values on every dataset and metric. Recall@20 improves over the strongest baseline by approximately 2.77\%, 4.81\%, 6.25\%, and 6.53\% on Baby, Sports, Clothing, and Electronics, respectively. These results show that the full pipeline improves ranking quality across the four evaluated domains. Compared with recent graph-based multimodal baselines, the gains are consistent with the motivation for transforming item geometry before propagation. The overall table alone cannot identify which design choices cause the improvements; the ablations below separate their contributions. Because standard top-$K$ metrics do not isolate complementarity, we next examine semantically distant, behaviorally connected cases. Similarity-island statistics, long-tail results, and qualitative analyses remain in Appendices~\ref{app:similarity_islands}, \ref{app:longtail}, and~\ref{app:visualization}.

\begin{table*}[t]
  \centering
  \caption{Performance comparison of baselines and MAGE in terms of Recall@K (R@K) and NDCG@K (N@K). The best results are \textbf{bolded}, and the second-best are \underline{underlined}. The improvements of MAGE over the strongest baseline are statistically significant under a paired t-test over five random seeds ($p<0.01$).}
  \label{tab:main-results}
  \setlength{\tabcolsep}{4pt}
  \resizebox{\textwidth}{!}{
  \begin{tabular}{llcccccccccc}
    \toprule
    \multirow{2}{*}{\textbf{Dataset}} & \multirow{2}{*}{\textbf{Metric}} 
    & \multicolumn{10}{c}{\textbf{Methods}} \\
    \cmidrule(lr){3-12}
    & & \textbf{MF-BPR} & \textbf{LightGCN} & \textbf{VBPR} & \textbf{SLMRec} & \textbf{BM3} & \textbf{MGCN} & \textbf{LGMRec} & \textbf{MENTOR} & \textbf{GUME} & \textbf{MAGE} \\
    \midrule
    \multirow{4}{*}{Baby}
      & R@10 & 0.0357 & 0.0479 & 0.0423 & 0.0540 & 0.0564 & 0.0620 & 0.0639 & \underline{0.0678} & 0.0673 & \textbf{0.0702} \\
      & R@20 & 0.0575 & 0.0754 & 0.0663 & 0.0810 & 0.0883 & 0.0964 & 0.0989 & \underline{0.1048} & 0.1042 & \textbf{0.1077} \\
      & N@10 & 0.0192 & 0.0257 & 0.0227 & 0.0285 & 0.0321 & 0.0339 & 0.0337 & 0.0362 & \underline{0.0365} & \textbf{0.0379} \\
      & N@20 & 0.0249 & 0.0328 & 0.0284 & 0.0357 & 0.0383 & 0.0427 & 0.0430 & 0.0450 & \underline{0.0460} & \textbf{0.0475} \\
    \midrule
    \multirow{4}{*}{Sports}
      & R@10 & 0.0432 & 0.0569 & 0.0558 & 0.0676 & 0.0656 & 0.0729 & 0.0719 & 0.0763 & \underline{0.0778} & \textbf{0.0821} \\
      & R@20 & 0.0653 & 0.0806 & 0.0856 & 0.1076 & 0.1087 & 0.1106 & 0.1068 & 0.1139 & \underline{0.1165} & \textbf{0.1221} \\
      & N@10 & 0.0241 & 0.0313 & 0.0307 & 0.0410 & 0.0408 & 0.0397 & 0.0387 & 0.0409 & \underline{0.0427} & \textbf{0.0440} \\
      & N@20 & 0.0298 & 0.0387 & 0.0384 & 0.0462 & 0.0443 & 0.0496 & 0.0477 & 0.0511 & \underline{0.0527} & \textbf{0.0544} \\
    \midrule
    \multirow{4}{*}{Clothing}
      & R@10 & 0.0187 & 0.0340 & 0.0423 & 0.0406 & 0.0482 & 0.0641 & 0.0555 & 0.0668 & \underline{0.0703} & \textbf{0.0738} \\
      & R@20 & 0.0279 & 0.0526 & 0.0663 & 0.0710 & 0.0743 & 0.0945 & 0.0828 & 0.0989 & \underline{0.1024} & \textbf{0.1088} \\
      & N@10 & 0.0103 & 0.0188 & 0.0223 & 0.0285 & 0.0231 & 0.0347 & 0.0302 & 0.0360 & \underline{0.0384} & \textbf{0.0396} \\
      & N@20 & 0.0126 & 0.0236 & 0.0284 & 0.0357 & 0.0281 & 0.0428 & 0.0371 & 0.0441 & \underline{0.0466} & \textbf{0.0484} \\
    \midrule
    \multirow{4}{*}{Electronics}
      & R@10 & 0.0158 & 0.0190 & 0.0229 & 0.0351 & 0.0343 & 0.0339 & 0.0341 & 0.0347 & \underline{0.0355} & \textbf{0.0383} \\
      & R@20 & 0.0241 & 0.0320 & 0.0337 & 0.0549 & 0.0545 & 0.0520 & 0.0517 & 0.0540 & \underline{0.0551} & \textbf{0.0587} \\
      & N@10 & 0.0076 & 0.0094 & 0.0113 & 0.0180 & 0.0171 & 0.0181 & \underline{0.0183} & \underline{0.0183} & 0.0178 & \textbf{0.0200} \\
      & N@20 & 0.0100 & 0.0124 & 0.0140 & 0.0228 & 0.0229 & 0.0227 & 0.0228 & \underline{0.0232} & \underline{0.0232} & \textbf{0.0252} \\
    \bottomrule
  \end{tabular}}
\end{table*}

\subsection{Performance on the Hard Complement Subset}
\label{sec:hard_complement}
To test the intended setting more directly, we select test positives connected through metadata co-purchase relations to a user's training-history item, with low visual and textual similarity to the selected history anchor. Both similarities must fall in the bottom 30\% of the eligible pool; Appendix~\ref{app:hard_complement} specifies the construction. This is a targeted diagnostic because it uses the relation type that motivates MAGE, while the primary benchmark remains the full-test evaluation above.

Table~\ref{tab:hard_complement} shows improvements over GUME and MENTOR on all 16 dataset--metric combinations. Relative to the stronger of the two baselines for each metric, the mean improvement is 9.87\%, with dataset means of 6.08\%, 14.00\%, 11.71\%, and 7.69\% on Baby, Sports, Clothing, and Electronics. The results support the ability to rank behaviorally connected but semantically distant items; they do not by themselves isolate which component produces the gain.

\begin{table}[t]
\centering\small
\caption{Performance on the hard complement subset. The stronger of GUME and MENTOR is the reference for the reported 9.87\% mean relative gain. Subset construction is a targeted diagnostic, not an independent full-test benchmark.}
\label{tab:hard_complement}
\setlength{\tabcolsep}{6pt}
\begin{tabular}{llrrrr}
\toprule
Dataset&Method&R@10&R@20&N@10&N@20\\\midrule
\multirow{3}{*}{Baby}&GUME&0.1093&0.1591&0.0590&0.0720\\
&MENTOR&0.1093&0.1599&0.0558&0.0687\\
&MAGE&\textbf{0.1184}&\textbf{0.1676}&\textbf{0.0628}&\textbf{0.0754}\\\midrule
\multirow{3}{*}{Sports}&GUME&0.1751&0.2345&0.0998&0.1150\\
&MENTOR&0.1616&0.2167&0.0952&0.1093\\
&MAGE&\textbf{0.2030}&\textbf{0.2682}&\textbf{0.1127}&\textbf{0.1297}\\\midrule
\multirow{3}{*}{Clothing}&GUME&0.1596&0.2245&0.0832&0.0996\\
&MENTOR&0.1411&0.1954&0.0701&0.0836\\
&MAGE&\textbf{0.1762}&\textbf{0.2567}&\textbf{0.0913}&\textbf{0.1119}\\\midrule
\multirow{3}{*}{Electronics}&GUME&0.0720&0.1134&0.0367&0.0470\\
&MENTOR&0.0704&0.1092&0.0314&0.0425\\
&MAGE&\textbf{0.0762}&\textbf{0.1191}&\textbf{0.0408}&\textbf{0.0511}\\
\bottomrule
\end{tabular}
\end{table}

A separate training-frequency-based grouping, without semantic-distance selection, also shows higher tail Recall for MAGE than matched GUME on Sports and Clothing (Appendix~\ref{app:popularity_control}). This complements the targeted hard-subset analysis.

\subsection{Ablation Study}
\label{sec:attribution}

We evaluate four block-level removals in Table~\ref{tab:ablation}. Removing Phase I retains complementary propagation, while removing complementary topology retains Phase-I features. Both removals reduce the two reported metrics on all four datasets, supporting their cumulative contribution.

\begin{table}[t]
\centering
\footnotesize
\caption{Ablation results of MAGE.}
\label{tab:ablation}
\setlength{\tabcolsep}{2.5pt}
\resizebox{\textwidth}{!}{
\begin{tabular}{lcccccccc}
\toprule
\multirow{2}{*}{\textbf{Variants}} 
& \multicolumn{2}{c}{\textbf{Baby}} 
& \multicolumn{2}{c}{\textbf{Sports}} 
& \multicolumn{2}{c}{\textbf{Clothing}}
& \multicolumn{2}{c}{\textbf{Electronics}} \\
\cmidrule(lr){2-3} \cmidrule(lr){4-5} \cmidrule(lr){6-7} \cmidrule(lr){8-9}
& \textbf{R@20} & \textbf{N@20} 
& \textbf{R@20} & \textbf{N@20} 
& \textbf{R@20} & \textbf{N@20}
& \textbf{R@20} & \textbf{N@20} \\
\midrule
\textbf{MAGE}
& \textbf{0.1077} & \textbf{0.0475}
& \textbf{0.1221} & \textbf{0.0544}
& \textbf{0.1088} & \textbf{0.0484}
& \textbf{0.0587} & \textbf{0.0252} \\
w/o Phase~I
& 0.1045 & 0.0458
& 0.1198 & 0.0510
& 0.1046 & 0.0472
& 0.0570 & 0.0246 \\
w/o Asym. Prop.
& 0.1070 & 0.0473
& 0.1210 & 0.0521
& 0.1073 & 0.0480
& 0.0580 & 0.0245 \\
w/o Comp. Topology
& 0.1061 & 0.0467
& 0.1210 & 0.0520
& 0.1049 & 0.0469
& 0.0552 & 0.0223 \\
w/o Gate
& 0.1050 & 0.0465
& 0.1211 & 0.0528
& 0.1060 & 0.0475
& 0.0566 & 0.0231 \\
\bottomrule
\end{tabular}
}
\end{table}

\paragraph{Architecture-level controls.}
\textbf{w/o Phase I.} Removing offline augmentation reduces both metrics on all four datasets while retaining complementary topology, supporting a contribution from geometric preprocessing beyond adding behavioral edges.
\textbf{w/o Asym. Prop.} Symmetric dual-stream propagation also underperforms the complete model, supporting distinct modality roles rather than the use of two streams alone.
\textbf{w/o Comp. Topology.} Retaining augmented features but removing complementary edges lowers performance, especially on Electronics. Thus, feature augmentation and complementary topology contribute jointly within the evaluated pipeline.
\textbf{w/o Gate.} Replacing the self-calibrated gate with concatenation reduces performance, supporting adaptive integration of the visual and textual streams. These block-level comparisons motivate finer controls, rather than establishing that every internal operation is necessary.

\paragraph{Simpler residuals and individual safeguards.}
Table~\ref{tab:phase1_fine} compares Phase-I variants with the remaining configuration and validation-based checkpoint selection fixed. Complete Phase I leads all 40 dataset--variant--metric comparisons in the full table (Appendix~\ref{app:rebuttal_controls}). Mean residuals and hard-SVD truncation remain competitive but are consistently lower. The gap constraint has the largest average contribution among the three removed safeguards, and orthogonalization the smallest. The differences are descriptive and modest; no separate statistical significance or indispensability claim is made for each safeguard.
\begin{table}[t]
\centering\small
\caption{Fine-grained Phase-I controls. Remaining settings are matched within each dataset. Full four-metric results are in Table~\ref{tab:phase1_full}.}
\label{tab:phase1_fine}
\begin{tabular}{lrrrr}
\toprule
&\multicolumn{2}{c}{Sports}&\multicolumn{2}{c}{Clothing}\\
\cmidrule(lr){2-3}\cmidrule(lr){4-5}
Variant&R@20&N@20&R@20&N@20\\\midrule
Complete Phase I&0.1221&0.0544&0.1088&0.0484\\
Mean residual&0.1211&0.0540&0.1074&0.0478\\
Hard-SVD residual&0.1206&0.0537&0.1072&0.0477\\
Without orthogonalization&0.1218&0.0543&0.1084&0.0482\\
Without gap constraint&0.1204&0.0536&0.1067&0.0475\\
Without normalized-gap backtracking&0.1209&0.0539&0.1080&0.0480\\
\bottomrule
\end{tabular}
\end{table}
\paragraph{Text preprocessing and transfer.}
Passing original non-LLM text embeddings through the same MAGE architecture changes the four metrics by $-0.45\%$ to $-1.07\%$ on Baby and $-1.03\%$ to $-1.69\%$ on Clothing. Sports changes by $+0.25\%$ to $+1.85\%$, but uses longer validation patience to avoid premature stopping; it is a convergence diagnostic, not a strictly matched superiority claim (Table~\ref{tab:raw_text_control}). Moreover, replacing only baseline text features with LLM-derived features produces mixed results, whereas offline augmentation improves all 24 reported Sports/Clothing metrics across GUME, MGCN, and MENTOR (Table~\ref{tab:offline_aug_generality}). Mixed Baby transfer results limit generalization. Together, these controls support geometric preprocessing beyond text refinement alone, without establishing universal transfer or ruling out simpler alternatives.

\subsection{Robustness and Efficiency}
\label{sec:evidence_scope}
Dropping or source-degree-preserving rewiring of 30\% of eligible co-purchase edges changes Sports/Clothing metrics by at most 1.96\% in magnitude. Each perturbed graph is used to recompute Phase-I features and complementary topology and retrain the model; the interaction graph, similarity graph, and evaluation set remain fixed (Appendix~\ref{app:graph_robustness}). These tests assess the reliability of imperfect co-purchase evidence under the specified corruptions.

Once embeddings are available, Phase I takes 1.89/4.39/4.36 seconds on Baby/Sports/Clothing with peak memory below 0.4 GB. These costs exclude text extraction and embedding generation. The gate adds 12,448 parameters relative to same-input GUME; no LLM is invoked during training or inference. Appendix~\ref{app:complexity} gives the scaling analysis.

\section{Conclusion}
In this paper, we proposed MAGE, a multimodal recommendation framework that incorporates heterophilic complementary signals through offline manifold augmentation and online asymmetric graph propagation. Experiments on real-world datasets show that MAGE consistently improves over strong baselines, especially under sparse and long-tail settings. Additional analyses indicate that MAGE better organizes functionally related yet semantically distant items, highlighting the potential of manifold-aware representation learning for complement-aware multimodal recommendation.

\textbf{Limitations.}
Co-purchase edges are imperfect evidence, and geometric safeguards do not certify the semantic meaning of individual residuals. The tested 30\% corruptions do not establish robustness to arbitrary graph noise, extreme sparsity, or temporal distribution shift. The hard-complement subset is a targeted diagnostic; transfer is mixed on Baby, and the Sports raw-text control changes validation patience. The design combines existing operators, with modest cumulative gains from individual safeguards. Simpler alternatives and broader deployment evaluation remain open directions.

\paragraph{Societal considerations.}
Complement-aware ranking may improve discovery, but co-purchase signals can also reflect popularity and commercial exposure biases. Offline accuracy does not establish user benefit or fair exposure; deployment should monitor these effects and the reliability of extracted product claims.

\begin{ack}
This work was supported in part by the National Natural Science Foundation of China (No.~62572075), in part by the Beijing Natural Science Foundation (Nos.~L233034 and L253004), and in part by the Fundamental Research Funds for the Beijing University of Posts and Telecommunications (No.~2025TSQY01).\par
\end{ack}

\clearpage
\bibliographystyle{plainnat}
\bibliography{ref}

\clearpage
\appendix
\setcounter{topnumber}{4}
\setcounter{bottomnumber}{3}
\setcounter{totalnumber}{6}
\renewcommand{\topfraction}{0.9}
\renewcommand{\bottomfraction}{0.85}
\renewcommand{\textfraction}{0.08}
\renewcommand{\floatpagefraction}{0.8}

\section{Supplementary Method Details}
\subsection{Gap-Constrained Injection Derivation}\label{app:gap_derivation}

A major risk of residual enhancement is over-injection: while stronger perturbations may introduce complementary information, they may also collapse the local semantic separation between substitute-like and complement-like neighbors. To address this issue, we do not use a fixed global injection coefficient. Instead, we compute an \textbf{adaptive injection strength} from a local substitute--complement gap constraint.

Let
\begin{equation}
    \boldsymbol{\mu}_{\mathcal{S}_i}
    =
    \frac{1}{|\mathcal{S}_i|}
    \sum_{j\in\mathcal{S}_i}\mathbf{x}_{j,t},
    \qquad
    \boldsymbol{\mu}_{\mathcal{W}_i}
    =
    \frac{1}{|\mathcal{W}_i|}
    \sum_{j\in\mathcal{W}_i}\mathbf{x}_{j,t},
\end{equation}
and define the original local substitute--complement gap as
\begin{equation}
    g_i^{\text{before}}
    =
    \mathbf{x}_{i,t}^{\top}
    \left(
    \boldsymbol{\mu}_{\mathcal{S}_i}
    -
    \boldsymbol{\mu}_{\mathcal{W}_i}
    \right).
\end{equation}
For $g_i^{\text{before}}>0$, we define the target gap as $g_i^{\mathrm{target}}=\rho g_i^{\text{before}}$ and require the enhanced representation to preserve at least this fraction of the original margin:
\begin{equation}
    g_i^{\text{after}}
    \ge
    \rho\, g_i^{\text{before}},
\end{equation}
where $\rho\in(0,1)$ is the keep ratio. The derivation below assumes $g_i^{\text{before}}>0$.

Using the unnormalized injected representation
\begin{equation}
    \mathbf{z}_i(\alpha)
    =
    \mathbf{x}_{i,t} + \alpha \mathbf{r}_i,
\end{equation}
the unnormalized post-injection gap can be written as
\begin{equation}
    g_i^{\mathrm{unnorm}}
    =
    \mathbf{z}_i(\alpha)^\top
    \left(
    \boldsymbol{\mu}_{\mathcal{S}_i}
    -
    \boldsymbol{\mu}_{\mathcal{W}_i}
    \right)
    =
    g_i^{\text{before}} + \alpha d_i,
\end{equation}
where
\begin{equation}
    d_i
    =
    \mathbf{r}_i^\top
    \left(
    \boldsymbol{\mu}_{\mathcal{S}_i}
    -
    \boldsymbol{\mu}_{\mathcal{W}_i}
    \right)
\end{equation}
measures how the residual direction changes the local gap. If $d_i \ge 0$, the unnormalized gap does not decrease, so we allow the capped injection. If $d_i < 0$, the constraint gives $\alpha\leq(1-\rho)g_i^{\text{before}}/(-d_i)$. We use the numerically stabilized bound
\begin{equation}
    \alpha_i^{\mathrm{gap}}
    =
    \frac{(1-\rho)\, g_i^{\text{before}}}{-d_i + \varepsilon},
\end{equation}
where $\varepsilon$ is a small constant for numerical stability. In practice, we use
\begin{equation}
    \alpha_i
    =
    \min(\alpha_{\max}, \alpha_i^{\mathrm{gap}})
    \quad \text{if } d_i < 0,
    \qquad
    \alpha_i
    =
    \alpha_{\max}
    \quad \text{if } d_i \ge 0,
\end{equation}
with the fixed safety cap $\alpha_{\max}=0.35$ in the public implementation. For fixed $g_i^{\text{before}}>0$ and $d_i<0$, increasing $\rho$ makes this initial coefficient non-increasing: the gap bound tightens, but the coefficient remains $0.35$ while the cap is active. A smaller $\rho$ permits more gap reduction; it does not necessarily produce a larger actual update.

The final enhanced textual representation is then
\begin{equation}
    \tilde{\mathbf{x}}_{i,t}
    =
    \frac{\mathbf{x}_{i,t} + \alpha_i \mathbf{r}_i}
    {\|\mathbf{x}_{i,t} + \alpha_i \mathbf{r}_i\|_2}.
\end{equation}
Since the final representation is $\ell_2$-normalized for cosine-based downstream learning, the unnormalized bound alone does not ensure the final gap condition. We further verify the normalized gap $g_i^{\text{after}}=\tilde{\mathbf{x}}_{i,t}^{\top}(\boldsymbol{\mu}_{\mathcal S_i}-\boldsymbol{\mu}_{\mathcal W_i})$ and backtrack $\alpha_i$ whenever it falls below $g_i^{\mathrm{target}}$.

Phase~I differs from standard manifold augmentation in two aspects. First, complementarity is extracted \emph{relatively} from each seed's local geometry through the strong/weak partition, rather than by imposing a global notion of similarity. Second, the residual is injected under an explicit \emph{gap-preserving rule} with post-normalization verification, keeping complementary enhancement bounded by the local substitute--complement structure. As a result, Phase~I introduces complement-aware directions into the textual space while controlling the risk of semantic drift before online propagation.

\subsection{Self-Calibrated Fusion Details}
\label{app:fusion_details}

Let $\mathbf{h}_{i,v}$ and $\mathbf{h}_{i,t}$ denote the final-layer item representations from the visual and textual streams. Since these two streams contain both shared category semantics and modality-specific complementary cues, we adopt a disentangle-then-fuse strategy that separates common structure from complementary residuals before integration.

We first extract a shared coarse-grained prototype by attention:
\begin{equation}
    \alpha_v, \alpha_t
    =
    \mathrm{Softmax}
    \left(
    \phi_{attn}(\mathbf{h}_{i,v}) \parallel \phi_{attn}(\mathbf{h}_{i,t})
    \right),
    \qquad
    \mathbf{h}_{i,\mathrm{coarse}}
    =
    \alpha_v \mathbf{h}_{i,v}
    +
    \alpha_t \mathbf{h}_{i,t}.
\end{equation}
This prototype captures the category-level identity jointly supported by both modalities.

To isolate complementary information, we subtract the shared prototype from each modality and modulate the resulting residuals using the collaborative representation:
\begin{equation}
   \mathbf{r}_{i,m}
   =
   \sigma_{beh}^{(m)}(\mathbf{e}_{i,\mathrm{id}})
   \odot
   \left(\mathbf{h}_{i,m} - \mathbf{h}_{i,\mathrm{coarse}}\right),
   \qquad
   m \in \{v,t\},
\end{equation}
where $\mathbf{e}_{i,\mathrm{id}}$ is the collaborative ID representation after graph propagation.

We then refine these residuals through a bidirectional self-calibrated gate. For each modality $m$, the gate combines cross-modal suggestion from the counterpart modality $\bar{m}$ and self-modal calibration:
\begin{equation}
    \mathbf{g}_{i,m}
    =
    \sigma
    \left(
    \mathrm{MLP}_{\bar{m}\rightarrow m}(\mathbf{r}_{i,\bar{m}})
    +
    \mathbf{W}_{m\rightarrow m}\mathbf{r}_{i,m}
    \right),
    \qquad
    m \in \{v,t\},
\end{equation}
where $\bar{m}$ denotes the counterpart modality of $m$. Instead of directly masking the residuals, we use a residual-style refinement:
\begin{equation}
   \tilde{\mathbf{r}}_{i,m}
   =
   \mathbf{r}_{i,m}
   +
   \beta_g \left(\mathbf{r}_{i,m} \odot \mathbf{g}_{i,m}\right),
   \qquad
   m \in \{v,t\}.
\end{equation}
This residual update retains the modality-specific signal and scales it using learned cross-modal and self-modal gates.

The fused item representation is then obtained by combining the coarse-grained prototype with the two refined residuals:
\begin{equation}
    \mathbf{e}_{i,\mathrm{fused}}
    =
    \frac{1}{3}
    \left(
    \mathbf{h}_{i,\mathrm{coarse}}
    +
    \tilde{\mathbf{r}}_{i,v}
    +
    \tilde{\mathbf{r}}_{i,t}
    \right),
    \qquad
    \mathbf{e}_i
    =
    \mathbf{e}_{i,\mathrm{id}}
    +
    \mathbf{e}_{i,\mathrm{fused}}.
\end{equation}

For users, we do not assume raw visual or textual inputs. Instead, we derive modality-conditioned user views by propagating visual and textual item signals over the user--item interaction graph. The resulting history-induced views, denoted as $\mathbf{e}_{u,v}^{\mathrm{ext}}$ and $\mathbf{e}_{u,t}^{\mathrm{ext}}$, are fused with the same coarse/residual/gating mechanism to obtain $\mathbf{e}_{u,\mathrm{fused}}$. The final user representation is
\begin{equation}
    \mathbf{e}_u
    =
    \mathbf{e}_{u,\mathrm{id}}
    +
    \mathbf{e}_{u,\mathrm{fused}}.
\end{equation}
Thus, item fusion operates on multimodal item streams, while user fusion operates on modality-conditioned interaction histories.

\subsection{Auxiliary Objectives}\label{app:objectives}
To reduce modality-wise scale mismatch, we use a lightweight moment-matching term
\begin{equation}
    \mathcal{L}_\mathrm{align}
    =
    \left| \mu(\mathbf{E}_v) - \mu(\mathbf{E}_t) \right|
    +
    \left| \sigma^2(\mathbf{E}_v) - \sigma^2(\mathbf{E}_t) \right|,
\end{equation}
where $\mathbf{E}_v$ and $\mathbf{E}_t$ denote the visual and textual stream embeddings.

To stabilize complementary augmentation, we align the fused multimodal representations with the collaborative backbone:
\begin{equation}
    \mathcal{L}_{BM}
    =
    \mathcal{L}_{\mathrm{NCE}}(\mathbf{e}_{u,\mathrm{fused}}, \mathbf{e}_{u,\mathrm{id}})
    +
    \mathcal{L}_{\mathrm{NCE}}(\mathbf{e}_{i,\mathrm{fused}}, \mathbf{e}_{i,\mathrm{id}}).
\end{equation}

For users, we additionally define a history-induced dual-stream representation
\begin{equation}
    \mathbf{e}_{u,\mathrm{expl}}
    =
    \frac{1}{2}
    \left(
    \mathbf{e}_{u,v}^{\mathrm{ext}}
    +
    \mathbf{e}_{u,t}^{\mathrm{ext}}
    \right),
\end{equation}
and regularize the fused user representation against it:
\begin{equation}
    \mathcal{L}_{\mathrm{UM}}
    =
    \mathcal{L}_{\mathrm{NCE}}(\mathbf{e}_{u,\mathrm{fused}}, \mathbf{e}_{u,\mathrm{expl}})
    +
    \mathcal{L}_{\mathrm{NCE}}(\tilde{\mathbf{e}}_{u,\mathrm{fused}}^{(1)}, \tilde{\mathbf{e}}_{u,\mathrm{fused}}^{(2)})
    +
    \mathcal{L}_{\mathrm{NCE}}(\tilde{\mathbf{e}}_{u,\mathrm{expl}}^{(1)}, \tilde{\mathbf{e}}_{u,\mathrm{expl}}^{(2)}),
\end{equation}
where perturbed views are generated by
\begin{equation}
    \tilde{\mathbf{e}}
    =
    \mathbf{e}
    +
    \eta \, \mathrm{sign}(\mathbf{e}) \odot
    \frac{\boldsymbol{\xi}}{\|\boldsymbol{\xi}\|_2+\varepsilon}.
\end{equation}

\section{Supplementary Experimental Setup}

\subsection{Dataset Statistics}
\label{app:data_stats}

Table~\ref{tab:data-stats} summarizes the basic statistics of the four benchmark datasets used in our experiments, including the numbers of users, items, interactions, sparsity, and the average number of co-purchase neighbors per item.

\begin{table}[!htbp]
  \centering
  \caption{Statistics of the experimental datasets.}
  \label{tab:data-stats}
  \begin{tabular}{lccccc}
    \toprule
    Dataset & User & Item & Inter. & Sparsity & Avg. Co-Pur. \\
    \midrule
    Baby        & 19,445   & 7,050  & 160,792   & 99.88\% & 23.9 \\
    Sports      & 35,598   & 18,357 & 296,337   & 99.95\% & 14.0 \\
    Clothing    & 39,387   & 23,033 & 278,677   & 99.97\% & 13.7 \\
    Electronics & 192,403  & 63,001 & 1,689,188 & 99.99\% & 25.6 \\
    \bottomrule
  \end{tabular}
\end{table}

\subsection{Implementation Details}
\label{app:implementation}
We implemented MAGE and baselines using the MMRec~\cite{zhou2023mmrec} framework on an NVIDIA RTX 3090 24GB GPU, fixing the embedding size to 64 and optimizing with Adam~\cite{DBLP:journals/corr/KingmaB14} using a learning rate of $10^{-3}$. In \textbf{Phase~I}, we construct the strong and weak neighbor sets using a per-seed relative percentile partition on local cosine similarities, and further apply a minimum similarity filter to suppress noisy weak co-purchase neighbors. We use a Soft Conceptor aperture of $\gamma=12.0$. The residual injection coefficient is not tuned as a fixed global scalar; instead, it is adaptively determined by a local gap-preservation constraint and further bounded by a safety cap ($\alpha_{\max}=0.35$ in the public implementation). In \textbf{Phase~II}, the loss weights $\lambda_{\mathrm{align}}$ (\texttt{vt\_loss}) and $\lambda_{\mathrm{info}}$ (\texttt{um\_loss}) were tuned in $\{10^{-4}, \dots, 10^{-1}\}$. The contrastive temperature $\tau_{bm}$ was fixed at $0.2$ on all four datasets; $\tau_{um}$ was fixed at $0.1$ on Baby and Sports and $0.2$ on Clothing and Electronics. The GCN depths were set to 3 for user--item interaction and 1 for item--item propagation. Following our asymmetric design, we set a unified neighbor count $k=10$ for both similarity and complementary edges, with the complementary edge modulator fixed at $\omega_{\mathrm{co}}=0.4$.

\subsection{Frozen Text Preprocessing and Failure Handling}
\label{app:preprocessing_protocol}
Qwen3-32B-AWQ is queried once per item through a local OpenAI-compatible endpoint, using the fixed prompt and schema in Appendix~\ref{app:prompt}. The configuration uses temperature 0.1, maximum output length 900 tokens, request timeout 120 seconds, and 48 asynchronous workers. Neither top-$p$ nor a decoding seed is explicitly set; they follow serving-backend defaults. The generated descriptions are frozen before representation learning so downstream runs use the same features. Freezing outputs supports downstream reproducibility but does not establish generation invariance across backends.

Responses are requested in JSON-object mode and checked for the seven required top-level fields and their container/value types. Request or parsing failures within an attempt are retried once; a complete generation that still fails validation is regenerated once. Persistently invalid items are excluded from normal preprocessing output and recorded in a separate item-level JSONL file with identifier, failure reason, and truncated diagnostics. This statement concerns preprocessing output, not an unreported change to the evaluation population.

Processed records retain assertions and their evidence spans, rejected claims, model-reported violations, and failure records. To support reproduction, the planned release includes exact prompts, preprocessing and validation code, serving configuration, frozen descriptions, the failure list, embedding-generation configuration, and 512-dimensional text features.

\section{Empirical Evidence for Similarity Islands}
\label{app:similarity_islands}

A key motivation of MAGE is that complementary items are often not naturally close in the raw embedding space, even when they are behaviorally associated through co-purchase relations. To directly examine this phenomenon, we compare \textbf{substitute} item pairs and \textbf{complementary} item pairs in terms of cosine similarity, and visualize their distributions in the original text embedding space.

\subsection{Pairwise Similarity Statistics}
\label{app:similarity_stats}

Table~\ref{tab:similarity_islands_stats} reports the average cosine similarity of substitute and complementary item pairs under both textual and visual modalities. Across all datasets, substitute pairs are consistently much more similar than complementary pairs, with clear gaps in both modalities. For example, on \textit{Baby}, the average text similarity drops from $0.7524$ for substitutes to $0.3748$ for complements, while the average visual similarity drops from $0.5614$ to $0.2044$. Similar trends hold on \textit{Sports}, \textit{Clothing}, and \textit{Electronics}. These statistics provide direct empirical evidence that complementary items are systematically more dispersed in the raw multimodal space, rather than being naturally grouped with semantically similar substitutes.

\begin{table}[!htbp]
\centering
\caption{Pairwise cosine similarity of substitute and complementary item pairs. The reported gap is defined as the difference between the average substitute similarity and the average complementary similarity. Larger gaps indicate stronger ``Similarity Islands'' effects.}
\label{tab:similarity_islands_stats}
\renewcommand{\arraystretch}{1.12}
\setlength{\tabcolsep}{8pt}
\small
\begin{tabular}{llccc}
\toprule
\textbf{Dataset} & \textbf{Modality} & \textbf{Substitute ($\mu \pm \sigma$)} & \textbf{Complement ($\mu \pm \sigma$)} & \textbf{Gap} \\
\midrule
Baby     & Text   & $0.7524 \pm 0.1099$ & $0.3748 \pm 0.0806$ & $0.3776$ \\
Baby     & Visual & $0.5614 \pm 0.1350$ & $0.2044 \pm 0.0719$ & $0.3570$ \\
Sports   & Text   & $0.7288 \pm 0.1154$ & $0.3640 \pm 0.0848$ & $0.3648$ \\
Sports   & Visual & $0.5894 \pm 0.1672$ & $0.1885 \pm 0.0857$ & $0.4009$ \\
Clothing & Text   & $0.7319 \pm 0.1241$ & $0.4652 \pm 0.1109$ & $0.2667$ \\
Clothing & Visual & $0.6306 \pm 0.1640$ & $0.3087 \pm 0.1366$ & $0.3219$ \\
Elec     & Text   & $0.8754 \pm 0.0740$ & $0.6154 \pm 0.0985$ & $0.2600$ \\
Elec     & Visual & $0.5537 \pm 0.2487$ & $0.2014 \pm 0.1056$ & $0.3522$ \\
\bottomrule
\end{tabular}
\end{table}

\subsection{Text-Space Visualization}
\label{app:similarity_tsne}

\begin{figure}[!htbp]
    \centering
    \begin{subfigure}{0.46\textwidth}
        \centering
        \includegraphics[width=\linewidth]{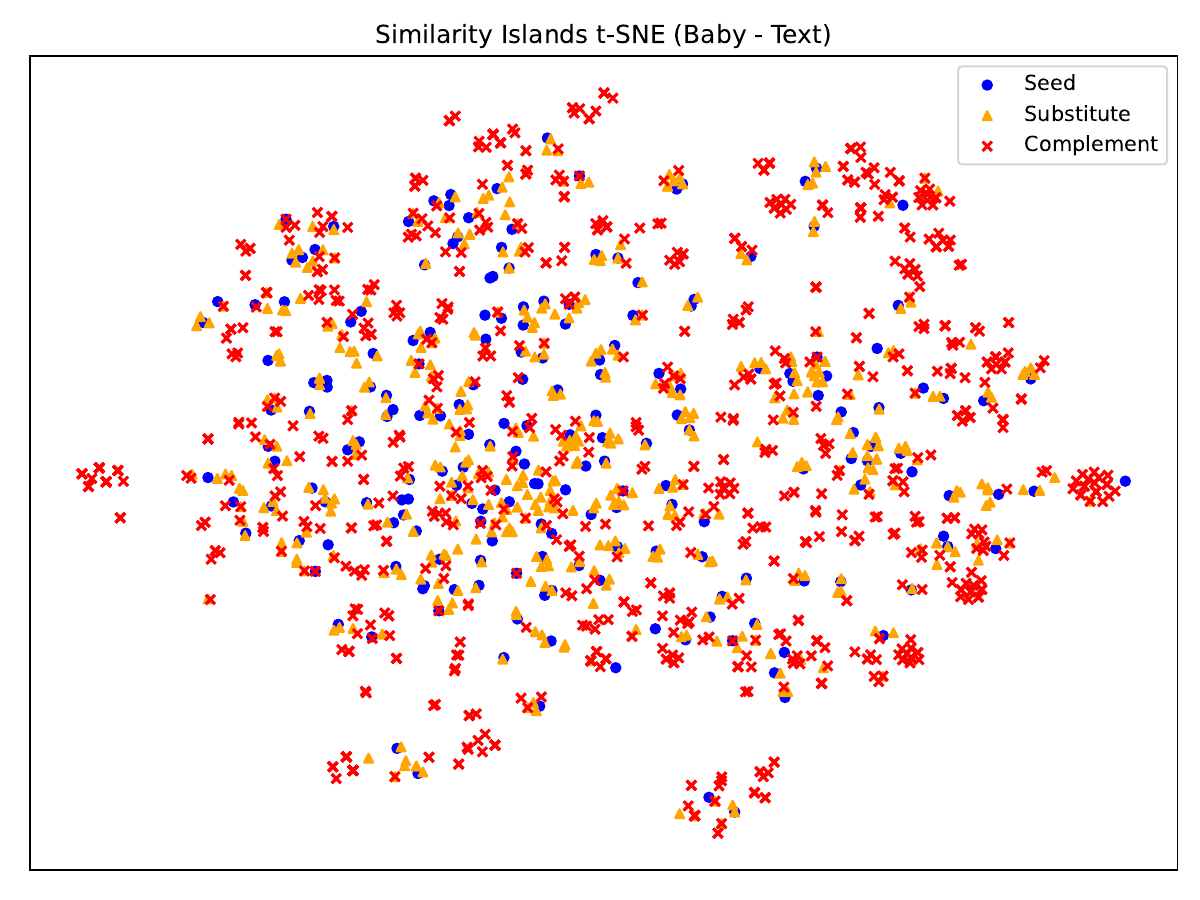}
        \caption{Baby}
        \label{fig:tsne_baby_text}
    \end{subfigure}
    \hfill
    \begin{subfigure}{0.46\textwidth}
        \centering
        \includegraphics[width=\linewidth]{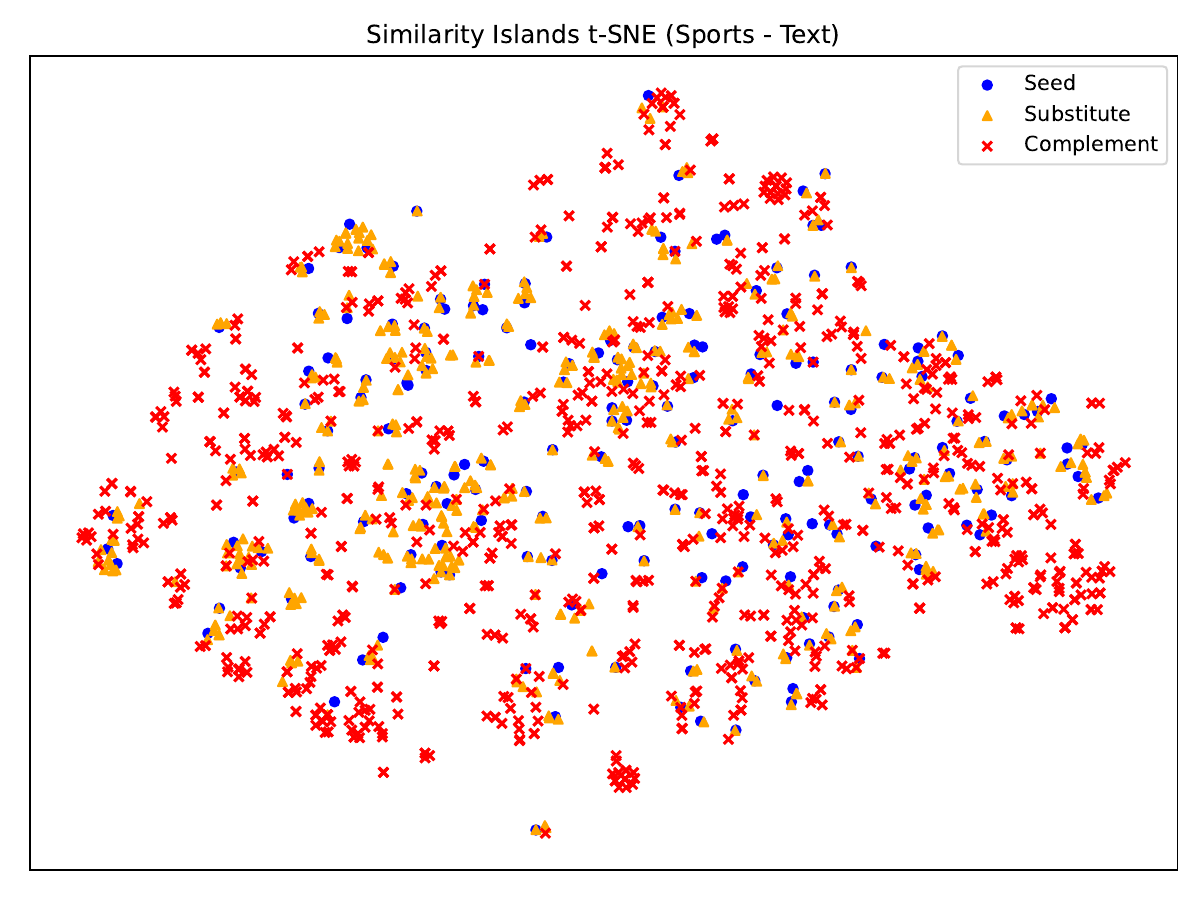}
        \caption{Sports}
        \label{fig:tsne_sports_text}
    \end{subfigure}

    \vspace{0.4em}

    \begin{subfigure}{0.46\textwidth}
        \centering
        \includegraphics[width=\linewidth]{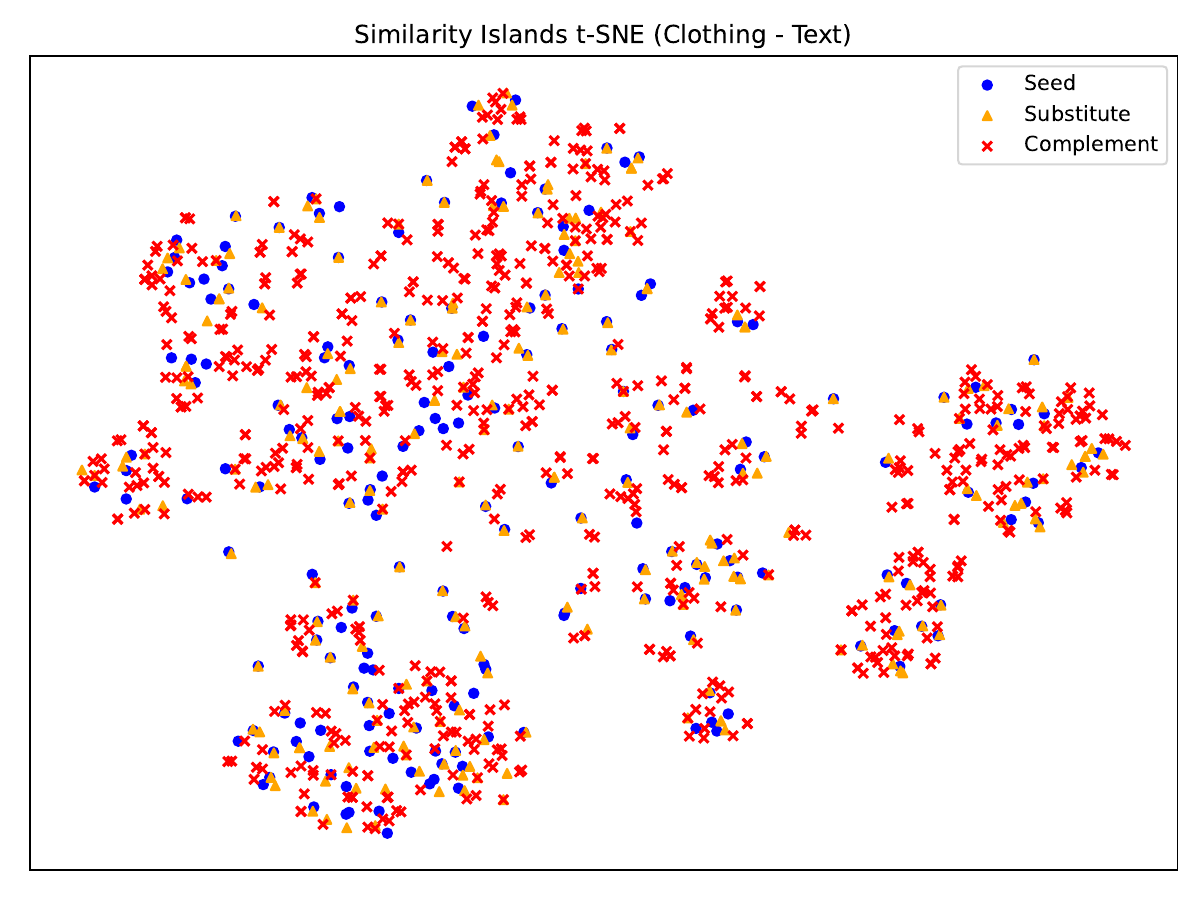}
        \caption{Clothing}
        \label{fig:tsne_clothing_text}
    \end{subfigure}
    \hfill
    \begin{subfigure}{0.46\textwidth}
        \centering
        \includegraphics[width=\linewidth]{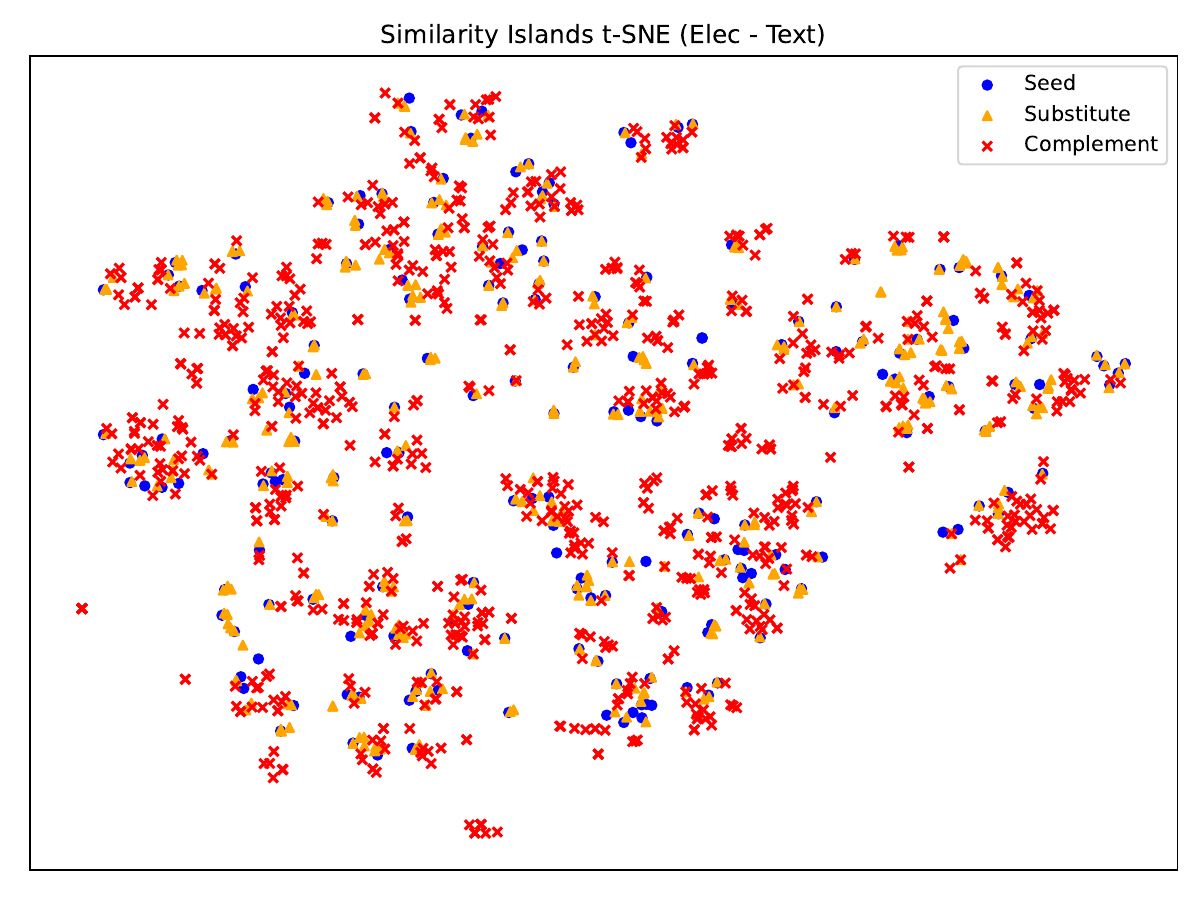}
        \caption{Electronics}
        \label{fig:tsne_elec_text}
    \end{subfigure}
    \caption{t-SNE visualization of the raw text embedding space for four datasets. Blue circles denote seed items, orange triangles denote substitutes, and red crosses denote complementary items. Across datasets, substitutes remain more locally concentrated around seed items, whereas complementary items are more dispersed, illustrating the ``Similarity Islands'' phenomenon.}
    \label{fig:similarity_islands_tsne}
\end{figure}

To complement the numerical statistics, Figure~\ref{fig:similarity_islands_tsne} visualizes the raw \emph{text} embedding space using t-SNE. In each plot, blue circles denote seed items, orange triangles denote substitutes, and red crosses denote complementary items. A consistent pattern can be observed across datasets: substitutes tend to remain closer to the seed items and occupy denser local neighborhoods, whereas complementary items are more scattered and often distributed across separate regions of the space. This effect is especially visible on \textit{Baby} and \textit{Sports}, and remains noticeable on \textit{Clothing} and \textit{Electronics} despite their higher textual similarity among complements.

Together, the statistics in Table~\ref{tab:similarity_islands_stats} and the visualization in Figure~\ref{fig:similarity_islands_tsne} support the core motivation of our method: complementary recommendation is not merely a matter of adding more edges in the graph, because the raw embedding space itself is organized around semantic similarity. This fragmented geometry forms what we call \textbf{Similarity Islands}, making it difficult for standard homophily-based propagation to recover complementary relations without introducing semantic drift.

\section{Generalization and Source of Performance Gains}
\label{sec:generality}\label{app:rebuttal_controls}

\subsection{Transferability Across Existing Multimodal Recommenders}
\label{sec:generality_transfer}

A natural question is whether the proposed offline manifold augmentation is intrinsically tied to MAGE, or can serve as a plug-and-play representation enhancement for existing multimodal recommenders. To answer this, we inject our augmented item embeddings into three strong baselines (\textbf{GUME}, \textbf{MGCN}, and \textbf{MENTOR}) without changing their architectures, graph construction, or training objectives. Concretely, we only replace the original item embeddings and keep all other settings fixed.

As shown in Table~\ref{tab:offline_aug_generality}, offline augmentation yields clear and consistent gains on \textit{Sports} and \textit{Clothing} across all three backbones, indicating that the proposed geometric enhancement provides a transferable inductive bias for modeling \textbf{heterophilic complementarity}. These improvements suggest that the benefit of our method is not restricted to the specific architecture of MAGE, but can generalize to existing multimodal recommenders as a lightweight plug-and-play enhancement.

On \textit{Baby}, however, the gains are marginal or occasionally negative, indicating that the benefit of transferring augmented features depends on the dataset and backbone. Differences in effective neighborhood coverage or co-purchase noise are possible explanations, but these factors were not isolated in this experiment. The results therefore do not identify sparsity as the cause of the mixed gains.

\subsection{LLM Text Refinement Alone Is Not Sufficient}
\label{sec:generality_llm}

Table~\ref{tab:offline_aug_generality} also allows us to separate the effect of \textbf{LLM-enhanced text embeddings} from that of \textbf{offline manifold augmentation}. Replacing only the textual feature file with LLM-enhanced embeddings does not produce consistent gains across models and datasets. Although \textbf{GUME} benefits slightly in some settings, the improvements are not stable across backbones; for \textbf{MGCN} and \textbf{MENTOR}, the LLM-text replacement is often marginal or even negative. This shows that better text-derived inputs alone are insufficient to resolve heterophily.

By contrast, \textbf{w/ Offline Aug} delivers clearer and more consistent improvements, especially on \textit{Sports} and \textit{Clothing}. This suggests that the main gain does not come from text refinement itself, but from explicitly reshaping the feature geometry so that semantically distant yet functionally complementary items become easier to connect during downstream graph propagation.

This observation is also consistent with the control design in Table~\ref{tab:offline_aug_generality}: feature replacement alone does not provide the same level of improvement as geometry-aware augmentation combined with graph propagation. What matters is the combination of complementary-aware geometric correction and graph-based propagation, rather than stronger text preprocessing by itself. We leave adaptive, noise-aware injection that automatically calibrates augmentation strength under extreme sparsity as future work.

\begin{table}[!htbp]
\centering
\caption{Generality analysis of offline manifold augmentation and LLM-enhanced text embeddings on existing multimodal recommenders. 
\textbf{Original} denotes the original backbone; \textbf{w/ LLM Text Emb} replaces only the textual feature file with LLM-enhanced embeddings; \textbf{w/ Offline Aug} replaces the original item embeddings with our offline manifold-augmented ones, while keeping all other settings unchanged.}
\label{tab:offline_aug_generality}
\renewcommand{\arraystretch}{1.08}
\setlength{\tabcolsep}{4pt}
\small
\resizebox{\textwidth}{!}{
\begin{tabular}{llcccc|cccc|cccc}
\toprule
\multirow{2}{*}{\textbf{Model}} & \multirow{2}{*}{\textbf{Setting}}
& \multicolumn{4}{c|}{\textbf{Baby}}
& \multicolumn{4}{c|}{\textbf{Sports}}
& \multicolumn{4}{c}{\textbf{Clothing}} \\
\cmidrule(lr){3-6} \cmidrule(lr){7-10} \cmidrule(lr){11-14}
& 
& \textbf{R@10} & \textbf{R@20} & \textbf{N@10} & \textbf{N@20}
& \textbf{R@10} & \textbf{R@20} & \textbf{N@10} & \textbf{N@20}
& \textbf{R@10} & \textbf{R@20} & \textbf{N@10} & \textbf{N@20} \\
\midrule

\multirow{3}{*}{\textbf{GUME}}
& Original
& 0.0673 & \textbf{0.1042} & 0.0365 & \textbf{0.0460}
& 0.0778 & 0.1165 & 0.0427 & 0.0527
& 0.0703 & 0.1024 & 0.0384 & 0.0466 \\
& w/ LLM Text Emb
& 0.0671 & 0.1015 & 0.0362 & 0.0451
& 0.0787 & 0.1176 & \textbf{0.0433} & \textbf{0.0537}
& 0.0707 & 0.1033 & 0.0385 & 0.0464 \\
& w/ Offline Aug
& \textbf{0.0674} & 0.1034 & \textbf{0.0366} & 0.0458
& \textbf{0.0796} & \textbf{0.1197} & \textbf{0.0433} & 0.0536
& \textbf{0.0712} & \textbf{0.1046} & \textbf{0.0387} & \textbf{0.0471} \\
\midrule

\multirow{3}{*}{\textbf{MGCN}}
& Original
& 0.0620 & \textbf{0.0964} & 0.0339 & \textbf{0.0427}
& 0.0729 & 0.1106 & 0.0397 & 0.0496
& 0.0641 & 0.0945 & 0.0347 & 0.0428 \\
& w/ LLM Text Emb
& 0.0611 & 0.0947 & 0.0334 & 0.0419
& 0.0722 & 0.1075 & 0.0396 & 0.0486
& 0.0624 & 0.0921 & 0.0335 & 0.0410 \\
& w/ Offline Aug
& \textbf{0.0621} & 0.0956 & \textbf{0.0340} & 0.0423
& \textbf{0.0764} & \textbf{0.1153} & \textbf{0.0417} & \textbf{0.0518}
& \textbf{0.0700} & \textbf{0.1018} & \textbf{0.0377} & \textbf{0.0458} \\
\midrule

\multirow{3}{*}{\textbf{MENTOR}}
& Original
& \textbf{0.0678} & \textbf{0.1048} & 0.0362 & 0.0450
& 0.0763 & 0.1139 & 0.0409 & 0.0511
& 0.0668 & 0.0989 & 0.0360 & 0.0441 \\
& w/ LLM Text Emb
& 0.0653 & 0.1029 & \textbf{0.0363} & \textbf{0.0460}
& 0.0756 & 0.1137 & 0.0409 & 0.0504
& 0.0685 & 0.1013 & 0.0372 & 0.0456 \\
& w/ Offline Aug
& 0.0671 & 0.1030 & 0.0361 & 0.0450
& \textbf{0.0787} & \textbf{0.1160} & \textbf{0.0426} & \textbf{0.0522}
& \textbf{0.0711} & \textbf{0.1044} & \textbf{0.0387} & \textbf{0.0471} \\
\bottomrule
\end{tabular}
}
\end{table}

\subsection{Fine-Grained Phase-I Ablation}
These controls use the submitted configuration, matched seed settings, and validation-selected checkpoints. They are reported descriptively; no separate significance claim is made for individual safeguards.
\begin{table}[!htbp]
\centering\small
\caption{Complete Phase-I controls from the discussion-stage experiments.}
\label{tab:phase1_full}
\begin{tabular}{llrrrr}
\toprule
Dataset&Variant&R@10&R@20&N@10&N@20\\\midrule
Clothing&Complete&0.0738&0.1088&0.0396&0.0484\\
&Mean residual&0.0729&0.1074&0.0391&0.0478\\
&Hard SVD&0.0727&0.1072&0.0390&0.0477\\
&Without orthogonalization&0.0735&0.1084&0.0395&0.0482\\
&Without gap constraint&0.0724&0.1067&0.0388&0.0475\\
&Without backtracking&0.0732&0.1080&0.0393&0.0480\\\midrule
Sports&Complete&0.0821&0.1221&0.0440&0.0544\\
&Mean residual&0.0815&0.1211&0.0437&0.0540\\
&Hard SVD&0.0811&0.1206&0.0435&0.0537\\
&Without orthogonalization&0.0819&0.1218&0.0439&0.0543\\
&Without gap constraint&0.0809&0.1204&0.0434&0.0536\\
&Without backtracking&0.0813&0.1209&0.0436&0.0539\\
\bottomrule
\end{tabular}
\end{table}

\subsection{Original-Text Dependency Control}
\begin{table}[!htbp]
\centering\small
\caption{Relative change (\%) when MAGE uses original non-LLM text. Sports uses longer validation patience and is not a strictly matched comparison.}
\label{tab:raw_text_control}
\begin{tabular}{lrrrr}
\toprule
Dataset&R@10&R@20&N@10&N@20\\\midrule
Baby&$-0.73$&$-0.45$&$-1.07$&$-0.86$\\
Sports&$+1.63$&$+0.25$&$+1.85$&$+0.86$\\
Clothing&$-1.23$&$-1.09$&$-1.69$&$-1.03$\\
\bottomrule
\end{tabular}
\end{table}
The same Phase-I transformation and Phase-II architecture are applied to original text embeddings. Checkpoints remain validation-selected. This control tests dependence on extracted text; it does not establish equivalence between preprocessing choices or isolate convergence effects on Sports.

\section{Complementarity and Long-Tail Diagnostics}
\subsection{Hard Complement Subset Evaluation}
\label{app:hard_complement}

Although MAGE is designed for general user--item recommendation rather than pure complementary-item retrieval, its central motivation is to better exploit complementary signals when they are behaviorally useful but semantically distant. To directly examine this property, we construct a \textbf{hard complement subset} from the test set.

For each test positive pair $(u,j)$, we first collect the user's training-history items $\mathcal{H}_u$ and require that at least one item $i\in\mathcal{H}_u$ is connected to $j$ through a co-purchase relation. The co-purchase graph is built from the union of \textit{also\_bought} and \textit{bought\_together} relations and treated as undirected. Among all co-purchased history anchors, we select the most semantically distant one according to the average of textual and visual cosine similarities:
\begin{equation}
    i^{*}
    =
    \arg\min_{i\in\mathcal{H}_u\cap\mathcal{N}_{\mathrm{co}}(j)}
    \frac{1}{2}
    \left(
    \mathrm{sim}_t(i,j)
    +
    \mathrm{sim}_v(i,j)
    \right).
\end{equation}
We then retain only those eligible test positives whose textual and visual similarities to $i^{*}$ both fall into the bottom 30\% of the eligible pool. This subset therefore focuses on cases that are behaviorally connected through co-purchase but semantically distant in both modalities. Because it uses the relation type that motivates MAGE, it is a targeted diagnostic rather than a distribution-independent benchmark; the primary evaluation remains the original full-ranking test set.

The corresponding results are reported in the main text (Table~\ref{tab:hard_complement}). The subset is defined for diagnostic evaluation; held-out user--item labels are not used to construct the metadata co-purchase graph.

\subsection{Qualitative Analysis: Visualizing the Functional Manifold}
\label{app:visualization}

To qualitatively examine whether MAGE bridges the semantic gap in complementary recommendation, we visualize the item embedding space of the \textit{Sports} dataset using t-SNE~\cite{maaten2008visualizing}.

\begin{figure}[!htbp]
    \centering
    \begin{subfigure}{0.48\textwidth}
        \centering
        \includegraphics[width=\linewidth]{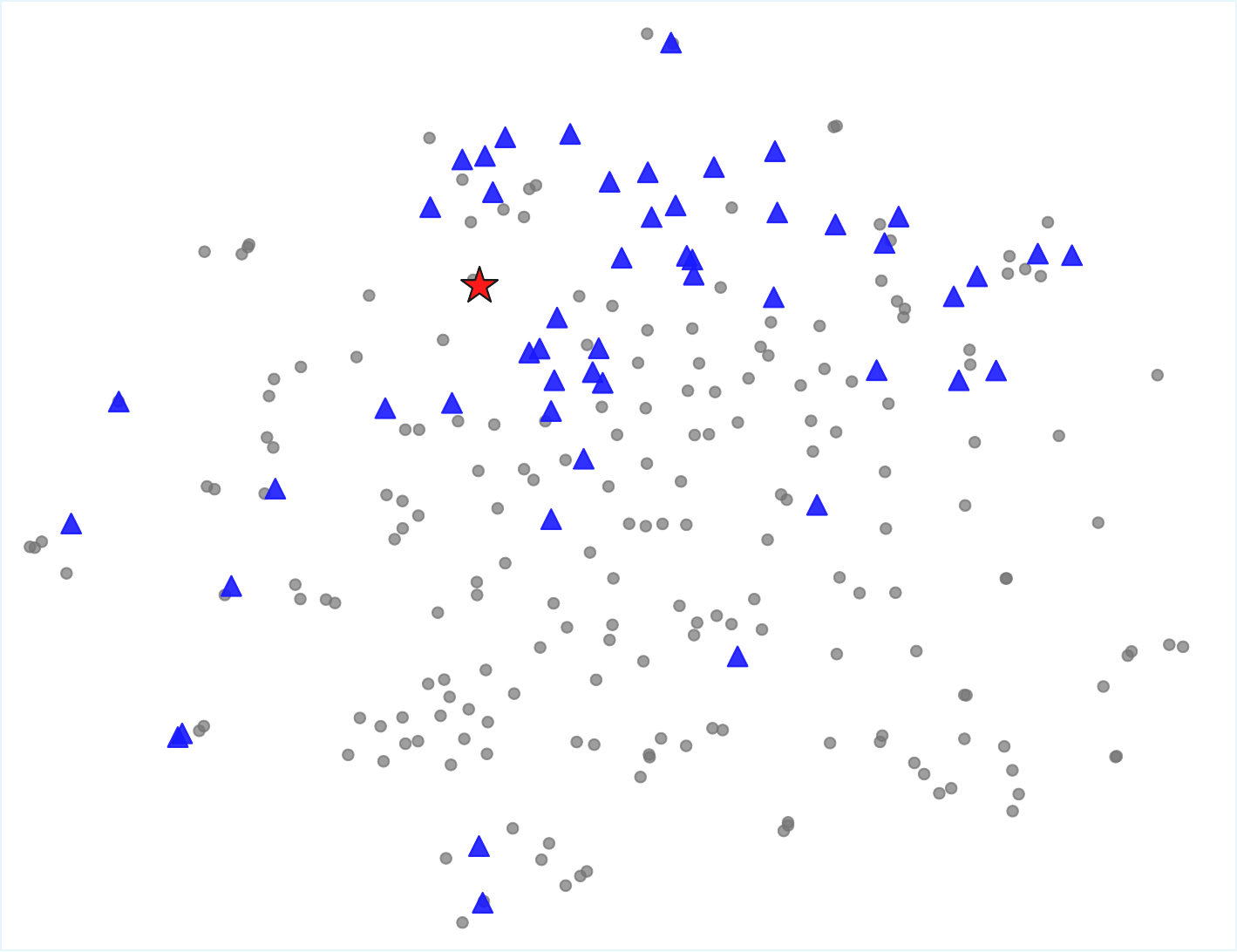}
        \caption{Baseline (GUME)}
        \label{fig:viz_gume_appendix}
    \end{subfigure}
    \hfill
    \begin{subfigure}{0.48\textwidth}
        \centering
        \includegraphics[width=\linewidth]{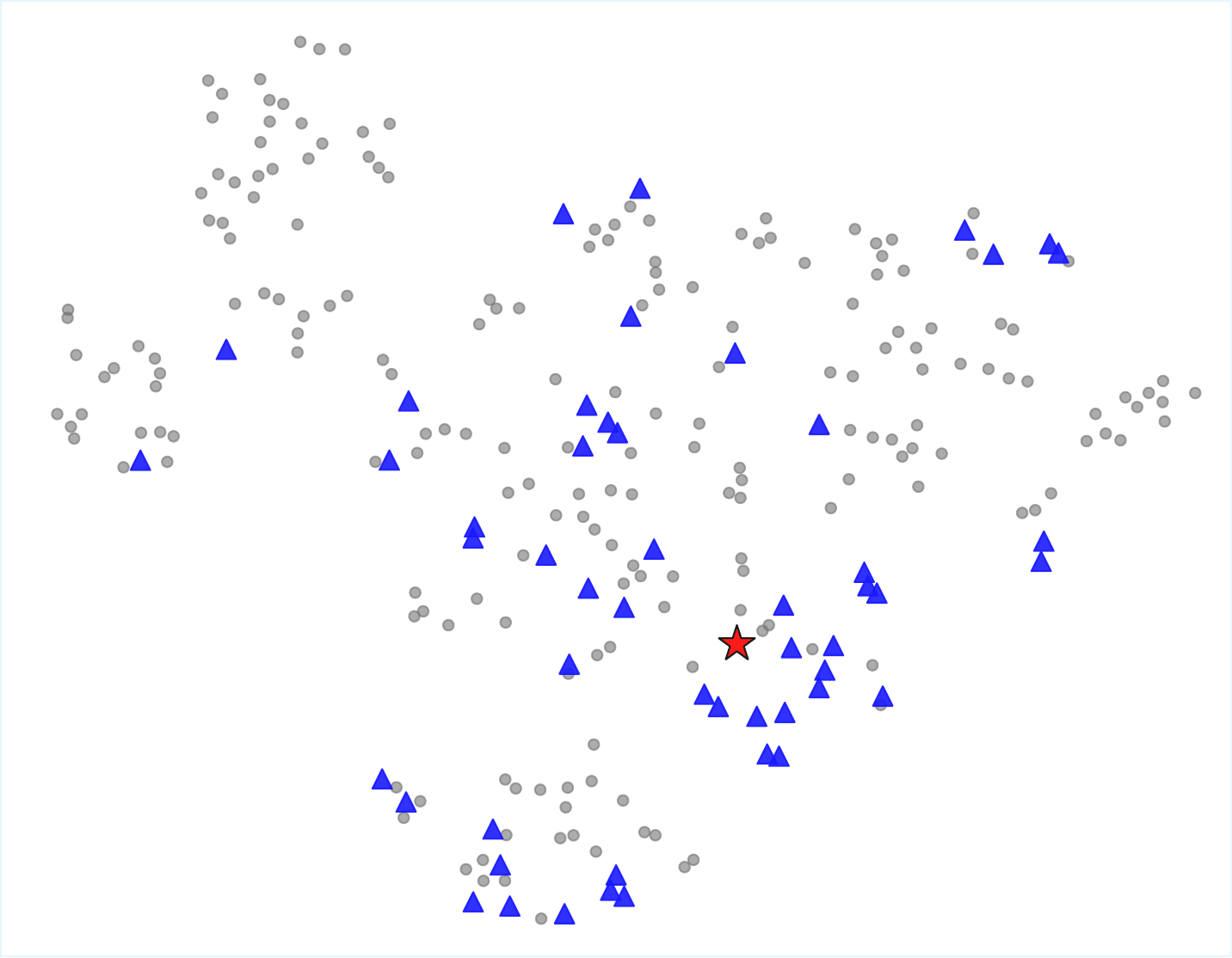}
        \caption{Ours (MAGE)}
        \label{fig:viz_mage_appendix}
    \end{subfigure}
    \caption{t-SNE visualization of item embeddings. \textbf{Red star}: target item (\textit{T-Tool}); \textbf{blue triangles}: functional complements (\textit{knives}); \textbf{gray dots}: semantic hard negatives (\textit{other tools}). Compared with the scattered distribution in GUME, MAGE aggregates functional complements into a more cohesive local neighborhood around the target item.}
    \label{fig:case_study_appendix}
\end{figure}

\paragraph{Setup.}
We select a \textit{Kershaw T-Tool} as the target item and consider two types of related items:
\textbf{functional complements} (co-purchased but semantically dissimilar, e.g., folding knives) and
\textbf{hard negatives} (textually similar but functionally unrelated, e.g., screwdrivers).
Figure~\ref{fig:case_study_appendix} compares the embedding distributions produced by GUME and MAGE.

\paragraph{Observations.}
In the selected visualization in Figure~\ref{fig:case_study_appendix}, the co-purchased items labeled as functional complements are more dispersed around the target under GUME.
Under MAGE, several of these items appear closer to the target, while the selected semantic hard negatives lie farther away.
This qualitative pattern is consistent with the intended local reorganization of item representations.
However, a selected two-dimensional projection does not independently verify functional compatibility or establish the same behavior across the full item population.

\subsubsection{Additional Geometric Case}
A supplementary case was selected using a fixed 75th-percentile threshold on Phase-I geometric gain, without inspecting product titles or test-ranking outcomes. For ``ILBE Main Pack USMC Generation 2 With Assault Pack,'' 9 of 10 observed top-10 co-purchase neighbors become closer after Phase I. Their median text-neighborhood rank improves from 3,872 to 1,441, and mean cosine similarity increases from 0.3442 to 0.4662. Selection on geometric gain makes this an illustration of the intended mechanism rather than an unbiased estimate of average improvement. The Kershaw T-Tool visualization in Appendix~\ref{app:visualization} provides a complementary qualitative example.

\subsection{Impact on Long-Tail Items}
\label{app:longtail}

\noindent
\begin{minipage}[t]{0.47\linewidth}
\vspace{0pt}
To examine MAGE's performance across item popularity levels, we performed a stratified evaluation on the Clothing dataset, following prior work~\cite{lin2024gume}. We partitioned items into five groups based on interaction frequency, where Group~1 represents the top 20\% (Head) and Groups~2--5 represent the long-tail (Tail), following the Pareto Principle~\cite{luo2023improving}. Figure~\ref{fig:longtail} reports the performance of MAGE compared with the strongest baselines, GUME and MENTOR.
\end{minipage}
\hfill
\begin{minipage}[t]{0.50\linewidth}
\vspace{0pt}
\centering
\includegraphics[width=\linewidth]{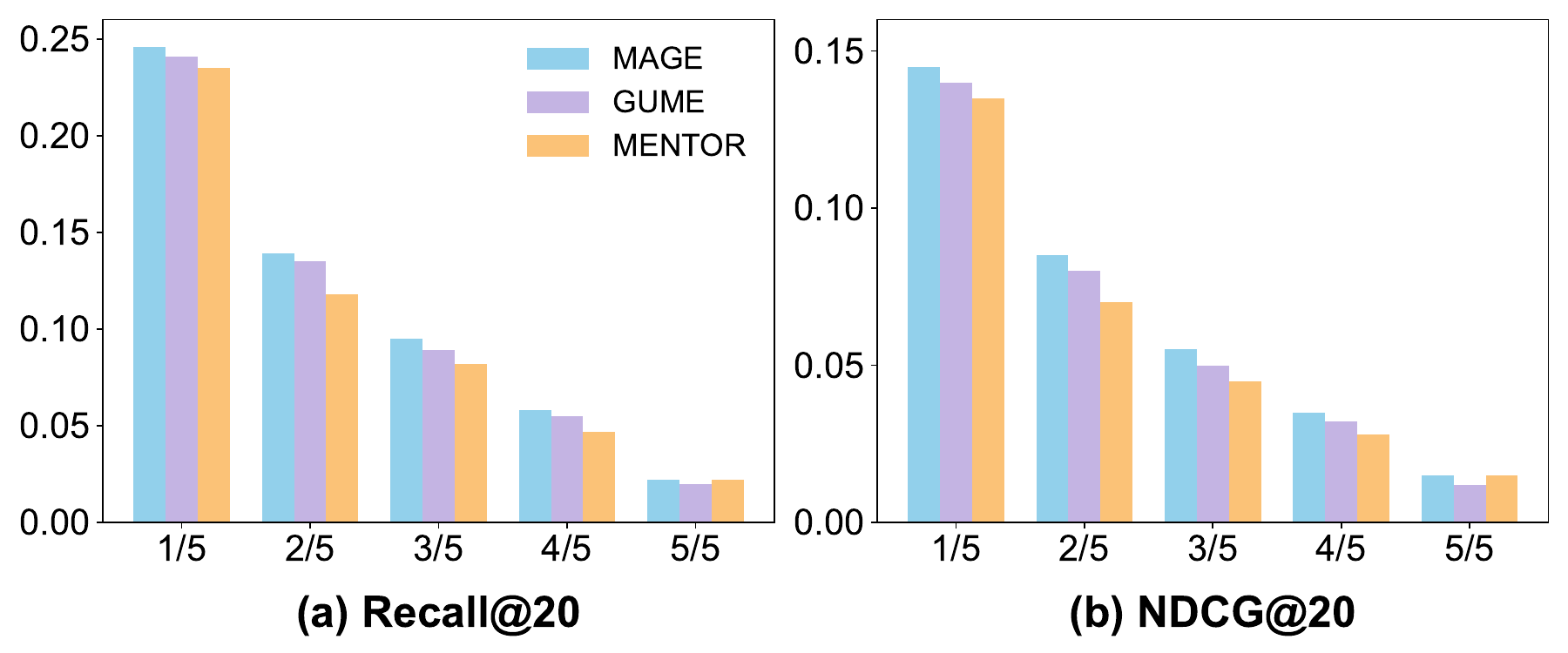}
\captionof{figure}{Performance comparison across item popularity groups on the Clothing dataset. Group~1 denotes Head items, while Groups~2--5 correspond to progressively sparser Tail items.}
\label{fig:longtail}
\end{minipage}

\vspace{0.5em}

The results show favorable ranking performance for MAGE across the displayed popularity groups, with gains in several tail groups as well as the Head group. These observations indicate that the measured benefits are not confined to popular items. They are consistent with the motivation to use co-purchase information for less frequently observed items, but do not isolate the mechanism responsible for the gains. In particular, group-level ranking metrics do not establish that popular-item representations or their relative ordering remain unchanged.

\subsubsection{Matched Popularity-Stratified Diagnostic}\label{app:popularity_control}
Head and tail groups are fixed using training interaction frequency, without semantic-distance filtering or changes to the full-ranking candidate set. Table~\ref{tab:tail_rebuttal} reports the available tail Recall measurements from this diagnostic, separately from the hard-complement subset.
\begin{table}[!htbp]
\centering\small
\caption{Tail Recall under the matched popularity-stratified evaluation.}
\label{tab:tail_rebuttal}
\begin{tabular}{llrr}
\toprule
Dataset&Method&R@10&R@20\\\midrule
Sports&GUME&0.0129&0.0282\\
&MAGE&0.0152&0.0320\\
Clothing&GUME&0.0161&0.0322\\
&MAGE&0.0210&0.0430\\
\bottomrule
\end{tabular}
\end{table}

\section{Sensitivity, Robustness, and Efficiency}
\label{app:hyper_complexity}
\subsection{Hyper-Parameter Sensitivity Analysis}
\label{app:hyperparams}

We further examine the sensitivity of MAGE with respect to four groups of hyper-parameters:
(\textit{i}) the Soft Conceptor aperture $\gamma$, which controls the strength of substitute-subspace filtering in Phase~I;
(\textit{ii}) the Phase-I keep ratio $\rho$, which determines how much local structural separation is preserved during the augmentation process;
(\textit{iii}) the auxiliary loss weights used in online optimization, namely the Info Loss and Align Loss coefficients; and
(\textit{iv}) the complementary neighborhood size $k$, which controls topological sparsity in Phase II.

\begin{figure}[!htbp]
    \centering
    \begin{subfigure}{0.47\textwidth}
        \centering
        \includegraphics[width=\linewidth]{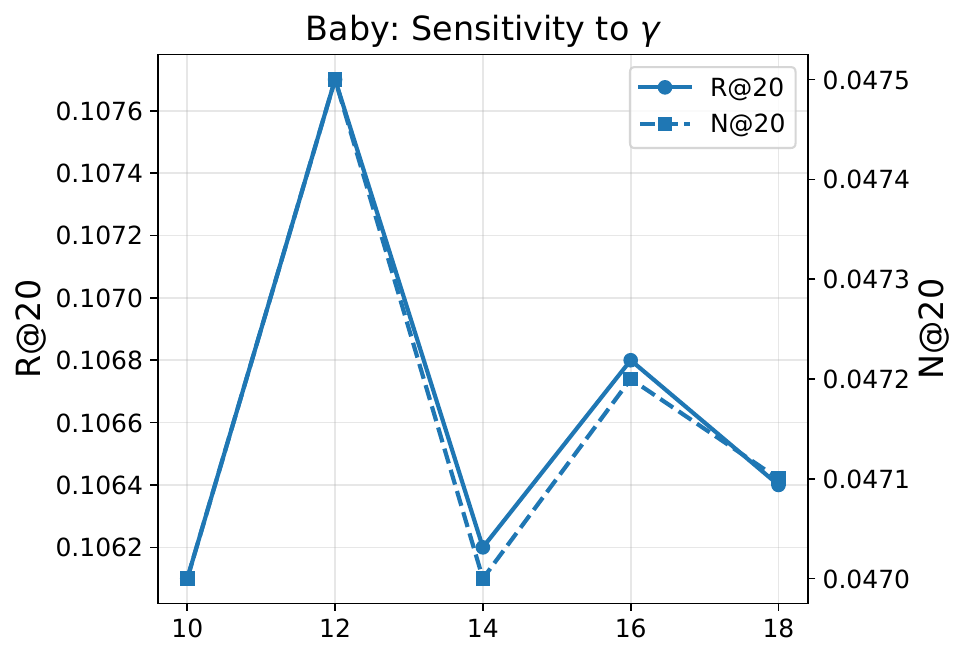}
        \caption{Baby}
        \label{fig:hyper_gamma_baby}
    \end{subfigure}
    \hfill
    \begin{subfigure}{0.47\textwidth}
        \centering
        \includegraphics[width=\linewidth]{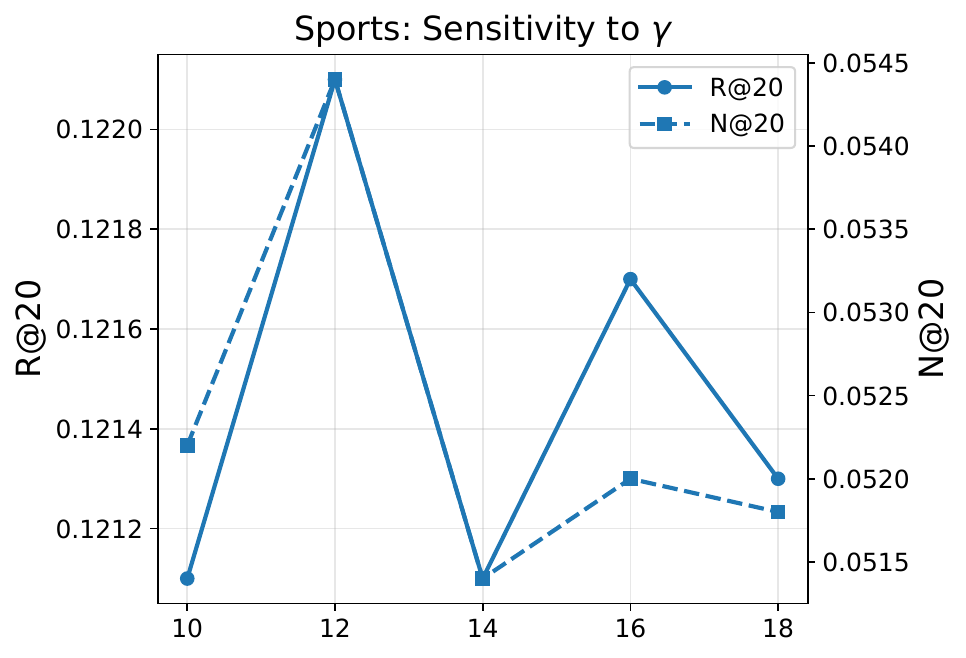}
        \caption{Sports}
        \label{fig:hyper_gamma_sports}
    \end{subfigure}

    \vspace{0.4em}

    \begin{subfigure}{0.47\textwidth}
        \centering
        \includegraphics[width=\linewidth]{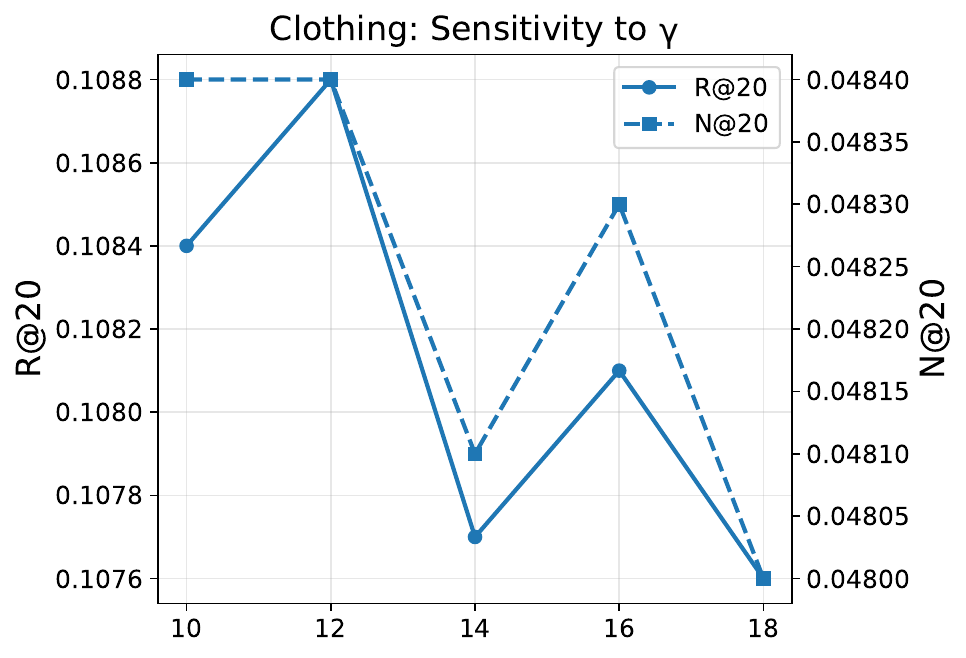}
        \caption{Clothing}
        \label{fig:hyper_gamma_clothing}
    \end{subfigure}
    \hfill
    \begin{subfigure}{0.47\textwidth}
        \centering
        \includegraphics[width=\linewidth]{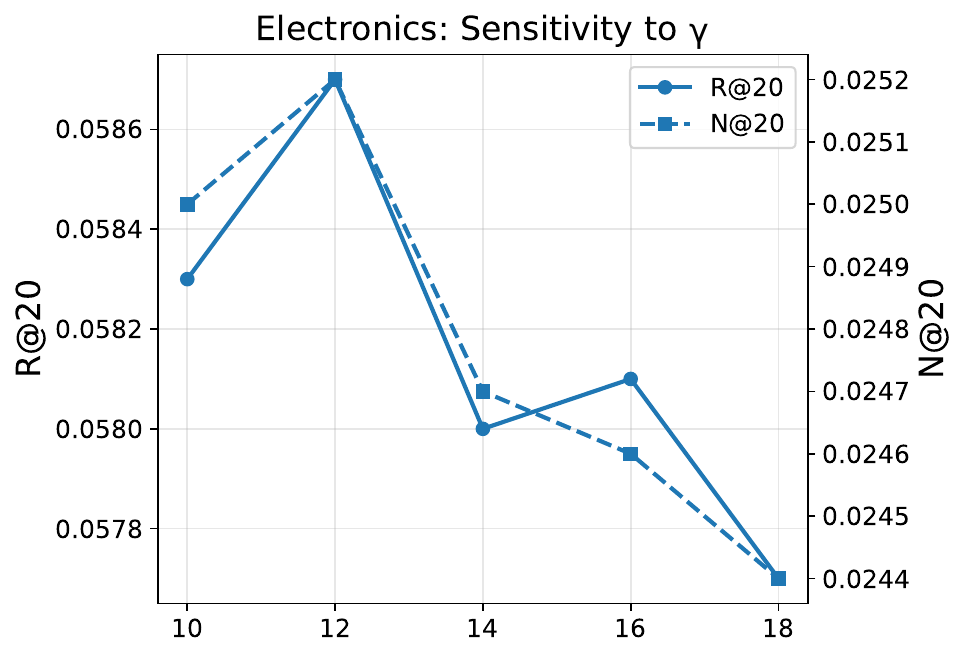}
        \caption{Electronics}
        \label{fig:hyper_gamma_elec}
    \end{subfigure}

    \caption{\textbf{Sensitivity to the Soft Conceptor aperture $\gamma$.}
    The best or near-best results occur in an intermediate regime. Baby R@20 and Sports N@20 at $\gamma=12$ are updated from Table~\ref{tab:main-results}; other points retain their original plotted values.}
    \label{fig:hyper_gamma}
\end{figure}

\paragraph{Sensitivity to the Soft Conceptor aperture $\gamma$.}
Figure~\ref{fig:hyper_gamma} shows the sensitivity of MAGE to the Soft Conceptor aperture $\gamma$ on all four datasets.
A consistent pattern emerges across Baby, Sports, Clothing, and Electronics:
performance is strongest in an intermediate regime, with $\gamma=12$ giving the best or near-best results in all cases.
When $\gamma$ is too small, the Soft Conceptor filtering is too weak to sufficiently suppress the substitute-dominated subspace.
When $\gamma$ becomes too large, the filtering becomes overly aggressive and may remove useful semantic information, leading to mild but consistent degradation.
This indicates that Phase~I benefits from \emph{moderate subspace filtering} rather than aggressive geometric perturbation.

\begin{figure}[!htbp]
    \centering
    \begin{subfigure}{0.47\textwidth}
        \centering
        \includegraphics[width=\linewidth]{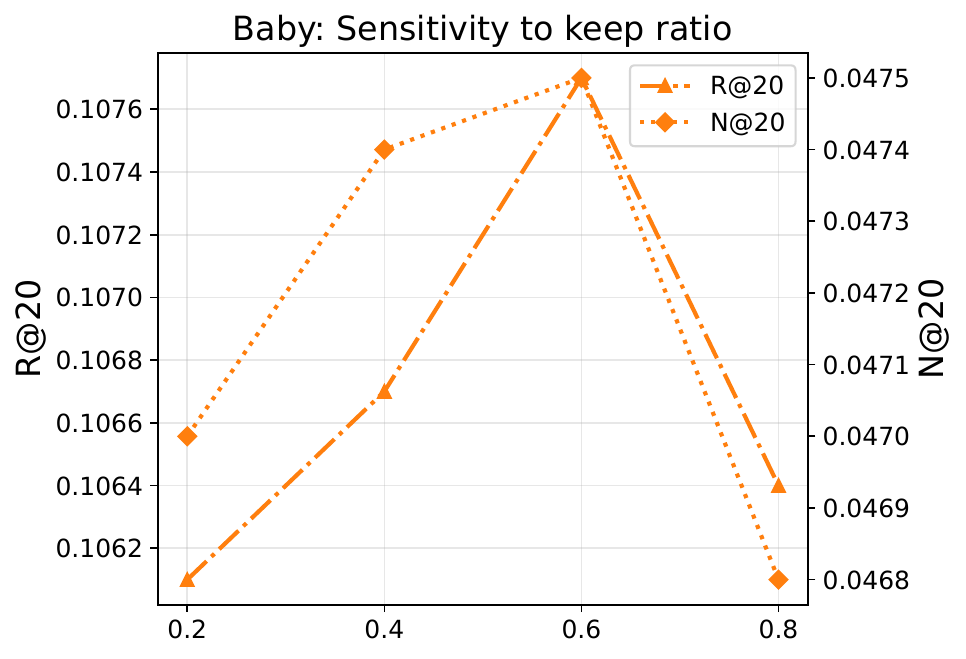}
        \caption{Baby}
        \label{fig:hyper_ratio_baby}
    \end{subfigure}
    \hfill
    \begin{subfigure}{0.47\textwidth}
        \centering
        \includegraphics[width=\linewidth]{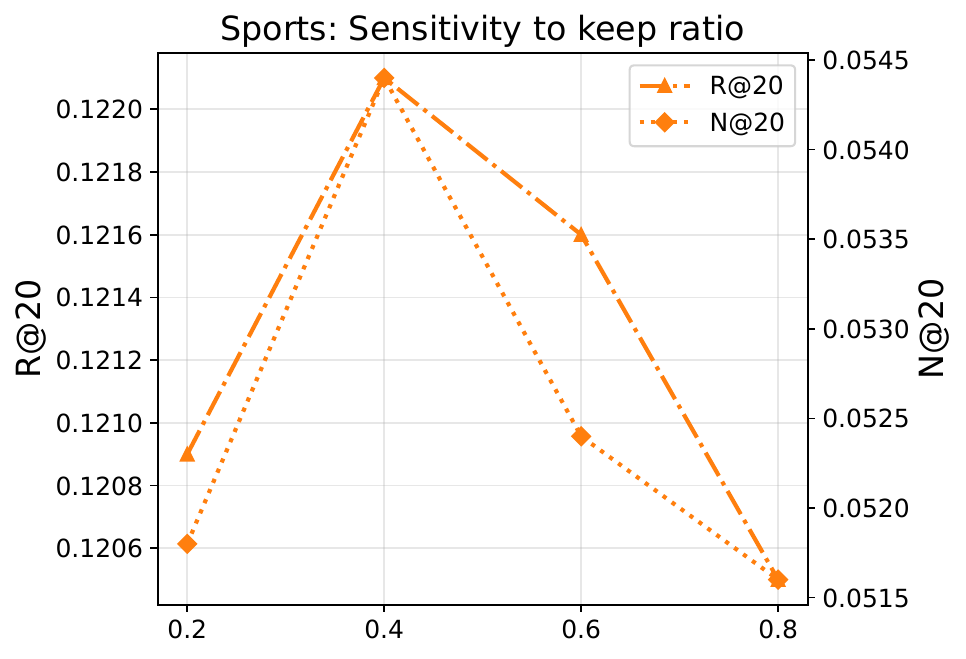}
        \caption{Sports}
        \label{fig:hyper_ratio_sports}
    \end{subfigure}

    \vspace{0.4em}

    \begin{subfigure}{0.47\textwidth}
        \centering
        \includegraphics[width=\linewidth]{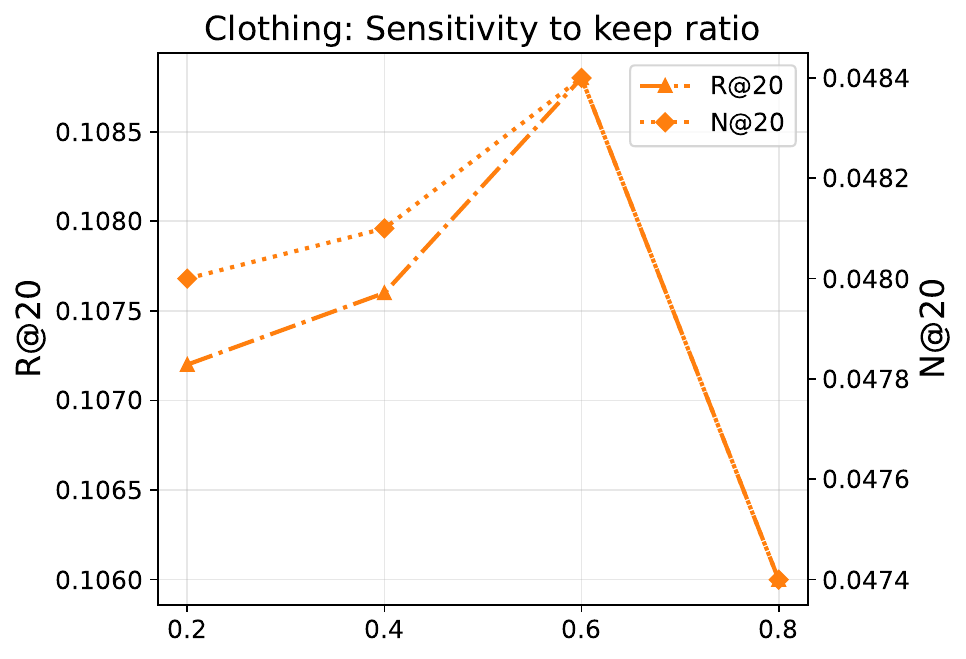}
        \caption{Clothing}
        \label{fig:hyper_ratio_clothing}
    \end{subfigure}
    \hfill
    \begin{subfigure}{0.47\textwidth}
        \centering
        \includegraphics[width=\linewidth]{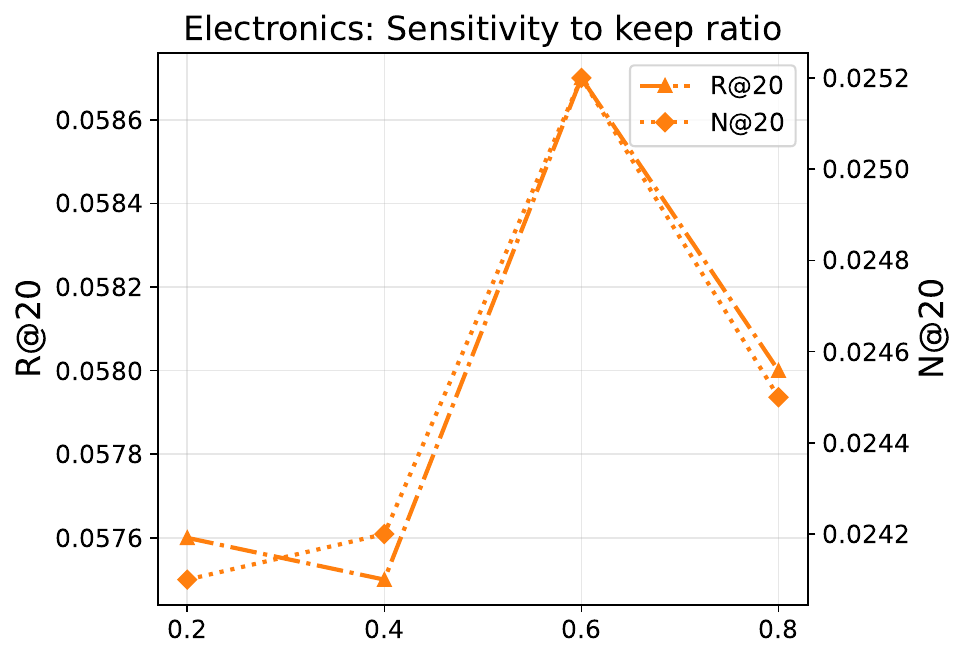}
        \caption{Electronics}
        \label{fig:hyper_ratio_elec}
    \end{subfigure}

    \caption{\textbf{Sensitivity to the Phase-I keep ratio $\rho$.}
    A larger keep ratio requires stricter local gap preservation. Baby R@20 at $\rho=0.6$ and Sports N@20 at $\rho=0.4$ are updated from Table~\ref{tab:main-results}; other points retain their original plotted values.}
    \label{fig:hyper_ratio}
\end{figure}

\paragraph{Sensitivity to the keep ratio $\rho$.}
Figure~\ref{fig:hyper_ratio} reports the effect of the keep ratio $\rho$, which controls how much of the local structural gap is preserved during offline augmentation.
In the displayed sweeps, both metrics peak at $\rho=0.4$ on Sports and at $\rho=0.6$ on Baby, Clothing, and Electronics.
A larger keep ratio requires stricter gap preservation and tightens the initial injection bound when the original gap is positive and the residual would reduce it; a smaller ratio relaxes this bound.
However, changing $\rho$ need not change the coefficient while the $0.35$ safety cap is active, and the normalized-gap check may further reduce the update.
The curves therefore show dataset-dependent performance under this constraint, rather than a monotonic relationship between $\rho$, injection magnitude, and ranking quality.

\begin{figure}[!htbp]
    \centering
    \begin{subfigure}{0.47\textwidth}
        \centering
        \includegraphics[width=\linewidth]{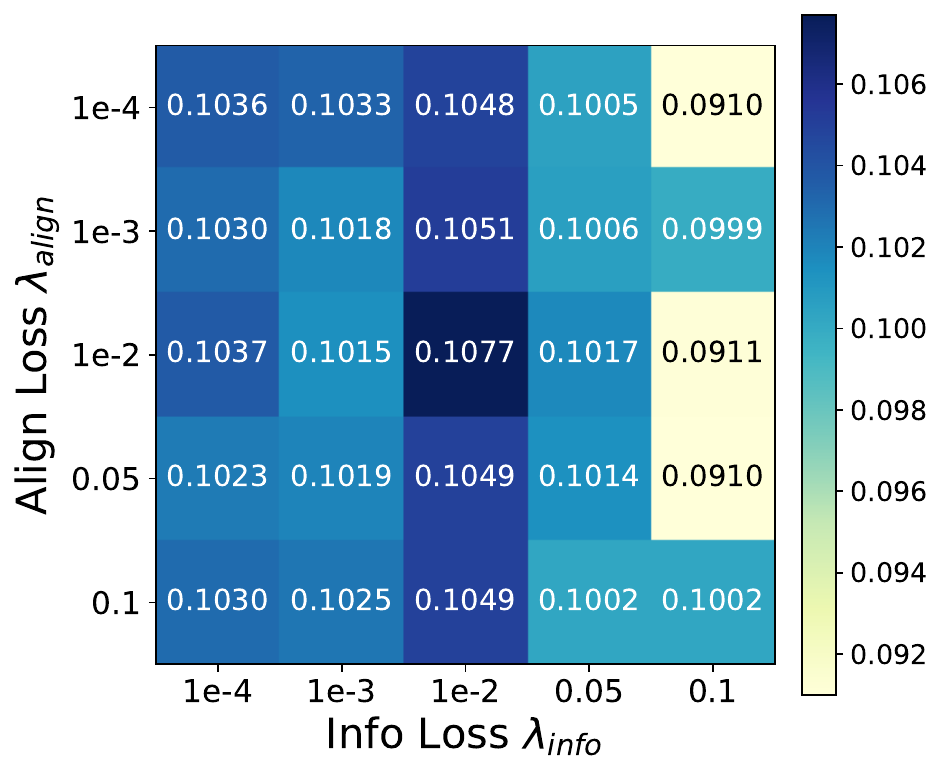}
        \caption{Loss weights: Baby}
        \label{fig:hyper_loss_baby}
    \end{subfigure}
    \hfill
    \begin{subfigure}{0.47\textwidth}
        \centering
        \includegraphics[width=\linewidth]{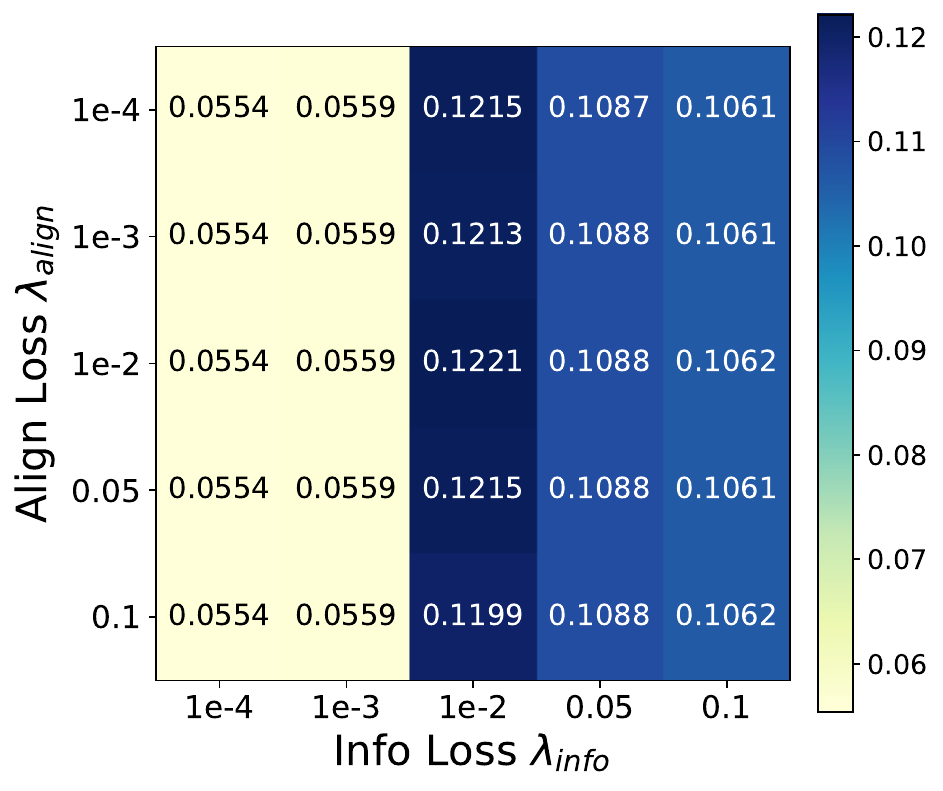}
        \caption{Loss weights: Sports}
        \label{fig:hyper_loss_sports}
    \end{subfigure}

    \vspace{0.4em}

    \begin{subfigure}{0.47\textwidth}
        \centering
        \includegraphics[width=\linewidth]{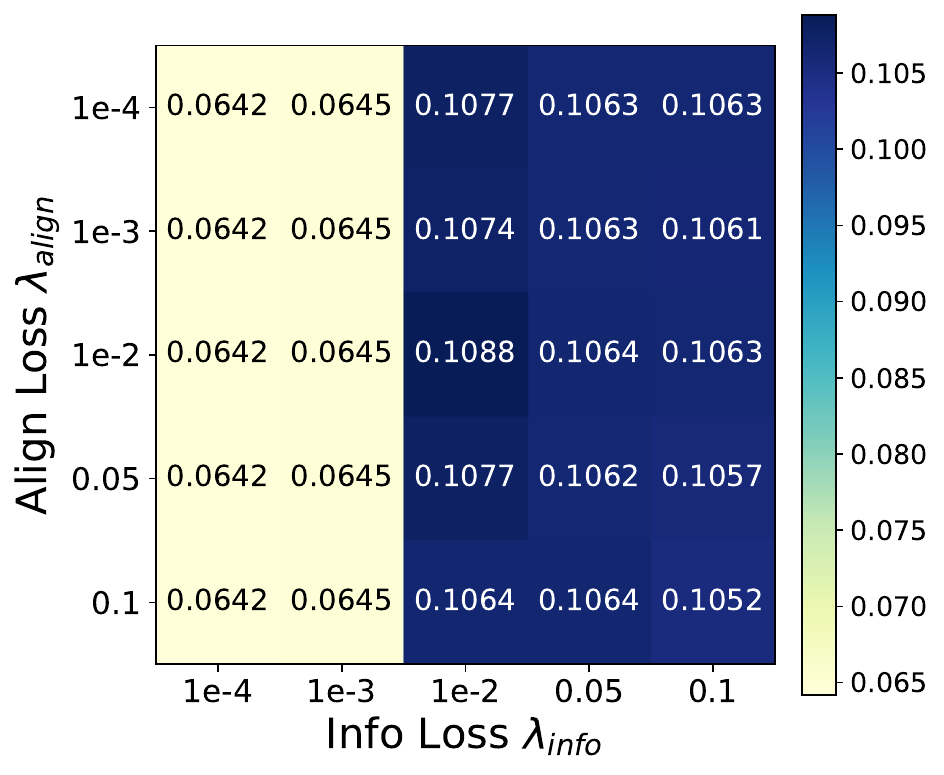}
        \caption{Loss weights: Clothing}
        \label{fig:hyper_loss_clothing}
    \end{subfigure}
    \hfill
    \begin{subfigure}{0.47\textwidth}
        \centering
        \includegraphics[width=\linewidth]{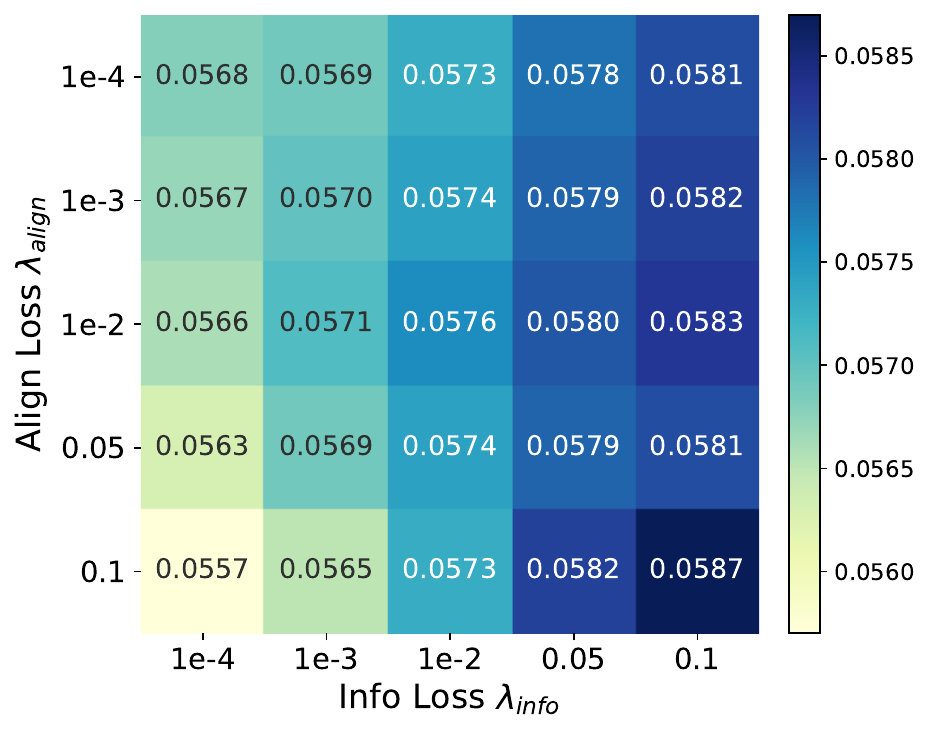}
        \caption{Loss weights: Electronics}
        \label{fig:hyper_loss_elec}
    \end{subfigure}

    \caption{\textbf{Sensitivity to online optimization.}
    Joint sensitivity of the Info Loss ($\lambda_{\mathrm{info}}$, UM) and Align Loss ($\lambda_{\mathrm{align}}$, VT) coefficients. Cell values and colors represent Recall@20. We fix $\tau_{bm}=0.2$ on all datasets and $\tau_{um}=0.1$ on Baby/Sports and $0.2$ on Clothing/Electronics.}
    \label{fig:hyper_online}
\end{figure}

\paragraph{Sensitivity to auxiliary loss weights.}
Figures~\ref{fig:hyper_loss_baby}--\ref{fig:hyper_loss_elec} report the interaction between the Info Loss and Align Loss coefficients on all four datasets. Here, the Info Loss coefficient $\lambda_{\mathrm{info}}$ weights $\mathcal{L}_{\mathrm{UM}}$ and corresponds to \texttt{um\_loss} in the implementation. The Align Loss coefficient $\lambda_{\mathrm{align}}$ weights $\mathcal{L}_{\mathrm{align}}$ and corresponds to \texttt{vt\_loss}. The temperature parameters are held at the dataset-specific values stated in the caption throughout these sweeps.
The best displayed results on Baby, Sports, and Clothing occur at $\lambda_{\mathrm{align}}=\lambda_{\mathrm{info}}=10^{-2}$.
Electronics differs: its highest displayed value occurs at $\lambda_{\mathrm{align}}=\lambda_{\mathrm{info}}=10^{-1}$, the upper boundary of the tested range.
The preferred regularization strength is therefore dataset-dependent; the sweeps do not support a universal claim that larger coefficients reduce performance.
The boundary optimum on Electronics also leaves behavior beyond the tested range unresolved. These observations motivate selecting the coefficients per dataset rather than imposing a single optimum across all four datasets.

\begin{figure}[!htbp]
    \centering
    \begin{subfigure}{0.52\textwidth}
        \centering
        \includegraphics[width=\linewidth]{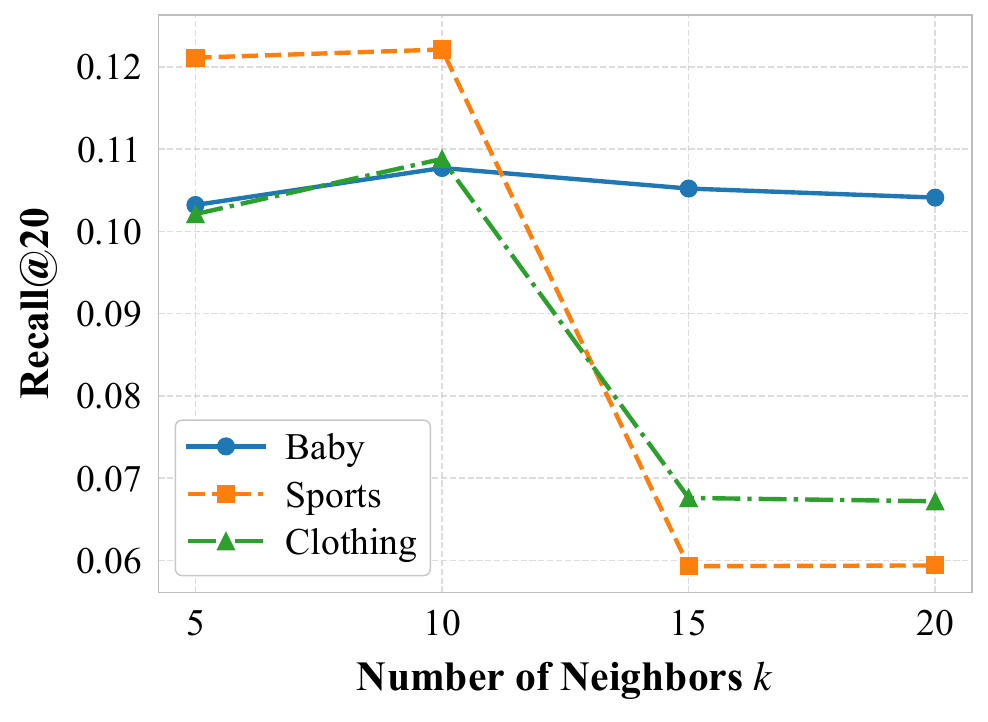}
        \caption{$k$-Sensitivity}
        \label{fig:hyper_k_sensitivity}
    \end{subfigure}

    \caption{\textbf{Sensitivity to topological sparsity in Phase II.}
    The complementary neighborhood size $k$ controls the sparsity of heterophilic propagation, and a relatively small neighborhood yields the most stable performance.}
    \label{fig:hyper_k_only}
\end{figure}

\paragraph{Sensitivity to topological sparsity in Phase~II.}
Figure~\ref{fig:hyper_k_sensitivity} reports the effect of the complementary neighborhood size $k$.
Despite differences in sparsity across datasets, the same qualitative trend is observed: performance peaks at a relatively small neighborhood and then declines as $k$ increases.
In particular, $k=10$ provides the most stable trade-off and remains consistently near-optimal.
This result highlights a key distinction between homophilic and heterophilic propagation:
while similarity graphs may benefit from denser smoothing, complementary graphs are much more noise-sensitive and therefore require sparse, precise propagation.

Taken together, these sweeps characterize the sensitivity of geometric augmentation, auxiliary regularization, and neighborhood size within the tested ranges.
They motivate dataset-specific choices of augmentation and regularization strengths, together with explicit gap control and neighborhood truncation.
The preferred settings depend on both the parameter and the dataset, as illustrated by the different loss-weight optima on Electronics. These results do not establish a single optimal configuration across datasets or beyond the evaluated ranges.

\subsection{Robustness to Co-purchase Graph Corruption}\label{app:graph_robustness}
With perturbation seed 999, we independently drop or rewire 30\% of eligible co-purchase edges; rewiring preserves source degree and total edge count. For each graph, we recompute Phase-I partitions and residuals, reconstruct complementary topology, and retrain from scratch with validation-only checkpoint selection. The interaction graph, similarity graph, and full-ranking evaluation remain fixed. Table~\ref{tab:graph_corruption} shows that relative metric changes remain within $\pm1.96\%$. This supports stability under the tested corruptions, not arbitrary noise or extreme sparsity.
\begin{table}[!htbp]
\centering\small
\caption{Relative change (\%) from the matched unperturbed configuration under 30\% graph corruption.}
\label{tab:graph_corruption}
\begin{tabular}{llrrrr}
\toprule
Dataset&Corruption&R@10&R@20&N@10&N@20\\\midrule
Sports&Dropout&$-0.25$&$+0.75$&$+1.72$&$+1.96$\\
Sports&Target rewiring&$+0.50$&$+1.25$&$+1.67$&$+1.51$\\
Clothing&Dropout&$-0.91$&$-0.83$&$-0.76$&$-0.41$\\
Clothing&Target rewiring&$-1.07$&$-0.79$&$-1.27$&$-1.03$\\
\bottomrule
\end{tabular}
\end{table}

\subsection{Complexity and Offline Efficiency}
\label{app:complexity}

Manifold-based methods are often criticized for poor scalability. MAGE remains efficient because both its geometric augmentation and graph propagation are localized.

\paragraph{Theoretical complexity.}
In Phase~I, offline manifold augmentation is performed independently for each item. For item $i$, SVD is applied only to a local matrix of size $(|\mathcal{S}_i|+1)\times d$, where $|\mathcal{S}_i|$ is bounded by a small neighborhood size and $d$ is fixed. The dominant per-item cost comes from local similarity computation, thin-SVD, and weak-neighbor projection:
\begin{equation}
    O(\bar{k}d + r^2d + |\mathcal{W}_i|rd),
\end{equation}
where $\bar{k}$ is the average co-purchase neighborhood size and $r\leq|\mathcal{S}_i|+1$ is the local SVD rank. Since these quantities are locally bounded in practice, the overall offline cost scales approximately linearly with the number of items $N$. This phase is executed once as preprocessing and introduces no overhead during training or inference.

In Phase~II, MAGE follows the standard efficiency profile of lightweight GCN-based recommenders: sparse graph convolution dominates the cost, while the self-calibrated gate only involves low-dimensional linear operations and element-wise products.

\paragraph{CPU-only benchmark.}
We further benchmark the Phase-I preprocessing on a CPU-only Apple Silicon machine using the three Amazon datasets. Phase-I preprocessing from existing embeddings, including its I/O but excluding LLM extraction and embedding generation, takes \textbf{1.89s}, \textbf{4.39s}, and \textbf{4.36s} on Baby, Sports, and Clothing, respectively, with peak memory below \textbf{0.4GB}. The median per-seed latency is only \textbf{0.16--0.20 ms}. A scalability test that clones the Baby graph from $1\times$ to $10\times$ shows near-linear growth, with per-seed latency increasing only from $0.170$ ms to $0.205$ ms. These results indicate that Phase~I is a one-shot, CPU-friendly preprocessing module, rather than a computational bottleneck.

\section{Prompt Design for Evidence-Bound Text Preprocessing}
\label{app:prompt}

To improve reproducibility, we provide the main prompt components used in our evidence-bound preprocessing pipeline. Consistent with the discussion in Section~\ref{sec:initialization}, the LLM is used only as a lightweight preprocessing tool to extract explicit functional attributes, normalize heterogeneous expressions, and remove marketing-heavy noise. It is not used to freely rewrite or semantically enrich the original text.

\subsection{System Instruction}

\begin{promptbox}{Prompt 1: System Instruction}
\begin{lstlisting}[style=promptstyle]
You are an evidence-bound information extractor and normalizer.

HARD RULES:
Only extract facts directly supported by verbatim snippets (<=12 words) from TITLE, DESCRIPTION, BRAND, or CATEGORIES; if no evidence, leave the field null or []. Do not invent or imply facts.

Use canonical product types {water_bottle, hydration_pack, backpack, sports_bag, yoga_mat, resistance_band, jump_rope, fitness_equipment, strength_training, recovery_tool, ball, goal_net, protective_gear, helmet, sport_eyewear, swim_gear, racquet_sport, golf_accessory, fishing_gear, cycling_accessory, wheel_sports, camping_gear, trekking_pole, lighting_outdoor, cooler, sports_towel, sports_socks, sportswear, dance_apparel, other}; assign based on clear textual cues, otherwise default to "other".

For potentially regulated items (e.g., "knife", "tactical knife", "spring assist", "pepper spray", "stun gun", "ammunition"):
- set product_type to "other"
- add a violation tag like "restricted_category: knife"
- keep copy strictly neutral (no marketing or purchase encouragement)

List fields must always be arrays. Do not assign age_range unless an exact numeric or stage is stated. Normalize units (oz -> ml; inch -> cm) when explicit; if a backpack/pack capacity in liters is explicit (e.g., "20L"), map to capacity_l (do not infer).

De-duplicate short values. Structured text must only use the extracted assertions, avoid marketing superlatives unless explicitly present, and maintain grammatical number consistency. Keep neutral tone; strictly neutral for regulated items.

STYLE:
- Keep outputs concise and concrete. No fluff.
- Avoid unverifiable attributes unless verbatim.
- Output VALID JSON only (no code fences).
\end{lstlisting}
\end{promptbox}

\subsection{Output Schema Constraints}

\begin{promptbox}{Prompt 2: Output Schema Constraints}
\begin{lstlisting}[style=promptstyle]
Output JSON with this schema:
{
  "normalized": {
    "product_type": "string|null",
    "materials": ["string", ...],
    "capacity_ml": number|null,
    "capacity_l": number|null,
    "age_range": "string|null",
    "compatibility": ["string", ...],
    "safety_claims": ["string", ...],
    "made_in": "string|null",
    "intended_use": ["string", ...]
  },
  "key_points": ["bullet 1", "bullet 2", "..."],
  "assertions": [
    {
      "claim": "string",
      "field": "product_type|materials|capacity_ml|capacity_l|age_range|compatibility|safety_claims|made_in|intended_use|other",
      "evidence": "verbatim snippet from input"
    }
  ],
  "structured_text": "string",
  "evidence": {
    "product_type": ["..."],
    "materials": ["..."],
    "capacity_ml": ["..."],
    "capacity_l": ["..."],
    "age_range": ["..."],
    "compatibility": ["..."],
    "safety_claims": ["..."],
    "made_in": ["..."],
    "intended_use": ["..."]
  },
  "rejected_claims": ["string", "..."],
  "violations": ["string", "..."]
}

CONSTRAINTS:
- If a field has no evidence, keep it null/[] and do NOT surface it in `assertions` or `structured_text`.
- Each assertion must quote its evidence verbatim (short phrases OK).
- If there is any uncertainty, prefer to omit and record it in `rejected_claims`.
- For regulated items (e.g., knives), do not include encouraging language; keep it purely descriptive using only evidence.
\end{lstlisting}
\end{promptbox}

\paragraph{Discussion.}
The two prompt components jointly enforce three properties required by our preprocessing pipeline: evidence grounding, schema-level normalization, and neutral noise filtering. In this way, the LLM serves only as a lightweight extractor/normalizer before feature encoding, rather than as a generator for open-ended text enhancement.

\section{Licenses and Terms of Use of Existing Assets}
\label{app:licenses}

Our experiments use several existing public assets, including benchmark datasets, open-source codebases, and pre-trained models.

\paragraph{Datasets.}
We use four public Amazon benchmark datasets (\textit{Baby}, \textit{Sports}, \textit{Clothing}, and \textit{Electronics}) obtained from the public collection released by McAuley and collaborators.\footnote{http://jmcauley.ucsd.edu/data/amazon/links.html}
We use these datasets only for academic research and follow their publicly stated access conditions and terms of use.

\paragraph{Code framework.}
Our implementation is built on the MMRec framework~\cite{zhou2023mmrec}. We use the public implementation released by its authors and follow the license and usage terms provided in the corresponding repository.

\paragraph{Pre-trained models.}
For evidence-bound preprocessing, we use Qwen3-32B-AWQ~\cite{yang2025qwen3}. This model is used only as a feature extractor or preprocessing tool, and we follow the corresponding license and terms of use provided by its official release.

\paragraph{License compliance.}
We credit the original creators of all existing assets used in this work and use them in accordance with their licenses and stated research-use conditions. We do not claim ownership of these external assets.

\section*{NeurIPS Paper Checklist}

\begin{enumerate}

\item {\bf Claims}
    \item[] Question: Do the main claims made in the abstract and introduction accurately reflect the paper's contributions and scope?
    \item[] Answer: \answerYes{} %
    \item[] Justification: The abstract and introduction clearly state the main motivation, methodological contributions, and empirical scope of the paper. These claims are consistent with the reported experimental results, including the overall performance comparison and the supporting analyses in the appendix.

\item {\bf Limitations}
    \item[] Question: Does the paper discuss the limitations of the work performed by the authors?
    \item[] Answer:  \answerYes{}
    \item[] Justification: The paper discusses limitations in the conclusion, including the dependence on co-purchase relation quality, possible degradation under noisy or extremely sparse co-purchase graphs, and the additional preprocessing/graph-construction costs introduced by the two-stage design.

\item {\bf Theory assumptions and proofs}
    \item[] Question: For each theoretical result, does the paper provide the full set of assumptions and a complete (and correct) proof?
    \item[] Answer: \answerNA{} %
    \item[] Justification: The paper does not present formal theoretical theorems or proofs. Our contributions are primarily methodological and empirical, focusing on architecture design, geometric augmentation, and experimental validation.

    \item {\bf Experimental result reproducibility}
    \item[] Question: Does the paper fully disclose all the information needed to reproduce the main experimental results of the paper to the extent that it affects the main claims and/or conclusions of the paper (regardless of whether the code and data are provided or not)?
    \item[] Answer: \answerYes{} %
    \item[] Justification: The paper describes the model architecture, optimization objective, datasets, feature preparation pipeline, evaluation protocol, and ablation settings in the main paper, with additional implementation details deferred to the appendix. We also provide an anonymized code repository to facilitate reproduction of the main results.

\item {\bf Open access to data and code}
    \item[] Question: Does the paper provide open access to the data and code, with sufficient instructions to faithfully reproduce the main experimental results, as described in supplemental material?
    \item[] Answer: \answerYes{} %
    \item[] Justification: An anonymized code repository is provided in the submission, and the datasets used in the paper are publicly available benchmark datasets. The supplemental material and appendix describe the experimental setup and implementation details needed to reproduce the reported results.

\item {\bf Experimental setting/details}
    \item[] Question: Does the paper specify all the training and test details (e.g., data splits, hyperparameters, how they were chosen, type of optimizer) necessary to understand the results?
    \item[] Answer: \answerYes{} %
    \item[] Justification: The paper specifies the datasets, feature preparation procedure, baseline methods, optimization objective, and evaluation metrics in the core text, while detailed implementation settings and hyper-parameter configurations are provided in the appendix. Together, these details are sufficient to understand and interpret the reported results.

\item {\bf Experiment statistical significance}
    \item[] Question: Does the paper report error bars suitably and correctly defined or other appropriate information about the statistical significance of the experiments?
    \item[] Answer: \answerYes{} %
    \item[] Justification: We report statistical significance information for the main experimental claims using repeated runs and paired significance testing. Specifically, the paper includes significance analysis over multiple random seeds for representative comparisons, and the corresponding testing protocol is described in the experimental discussion.

\item {\bf Experiments compute resources}
    \item[] Question: For each experiment, does the paper provide sufficient information on the computer resources (type of compute workers, memory, time of execution) needed to reproduce the experiments?
    \item[] Answer: \answerYes{} %
    \item[] Justification: The paper specifies the main compute environment used for experiments, including the GPU type and memory configuration, in the appendix implementation details. These resources are sufficient to understand the scale of the reported experiments and to reproduce the setup.
    
\item {\bf Code of ethics}
    \item[] Question: Does the research conducted in the paper conform, in every respect, with the NeurIPS Code of Ethics \url{https://neurips.cc/public/EthicsGuidelines}?
    \item[] Answer: \answerYes{} %
    \item[] Justification: To the best of our knowledge, the research described in this paper conforms to the NeurIPS Code of Ethics. The work uses public benchmark datasets, preserves anonymity at submission time, and does not involve human subjects or sensitive private data collection.

\item {\bf Broader impacts}
    \item[] Question: Does the paper discuss both potential positive societal impacts and negative societal impacts of the work performed?
    \item[] Answer: \answerYes{} %
    \item[] Justification: The paper discusses both potential benefits and risks of improved complementary recommendation. On the positive side, better multimodal recommendation may improve retrieval quality and user experience; on the negative side, such systems may amplify existing behavioral biases, create filter effects, or reinforce commercially skewed exposure if deployed without proper monitoring and safeguards.
    
\item {\bf Safeguards}
    \item[] Question: Does the paper describe safeguards that have been put in place for responsible release of data or models that have a high risk for misuse (e.g., pre-trained language models, image generators, or scraped datasets)?
    \item[] Answer: \answerNA{} %
    \item[] Justification: The paper does not release high-risk generative models, scraped sensitive datasets, or other assets that would require special misuse safeguards beyond standard anonymized code release and citation of public benchmark data sources.

\item {\bf Licenses for existing assets}
    \item[] Question: Are the creators or original owners of assets (e.g., code, data, models), used in the paper, properly credited and are the license and terms of use explicitly mentioned and properly respected?
    \item[] Answer: \answerYes{} %
    \item[] Justification: We properly credit the public datasets, code framework, and pre-trained model used in this work, and we explicitly state their sources and licenses or terms of use in the appendix. The external model is used only for preprocessing or feature extraction under its official usage conditions.

\item {\bf New assets}
    \item[] Question: Are new assets introduced in the paper well documented and is the documentation provided alongside the assets?
    \item[] Answer: \answerNA{} %
    \item[] Justification: The paper does not introduce a new dataset, model family, or benchmark as a standalone released asset. The main contribution is a new recommendation method evaluated on existing public datasets.

\item {\bf Crowdsourcing and research with human subjects}
    \item[] Question: For crowdsourcing experiments and research with human subjects, does the paper include the full text of instructions given to participants and screenshots, if applicable, as well as details about compensation (if any)? 
    \item[] Answer: \answerNA{} %
    \item[] Justification: The paper does not involve crowdsourcing experiments or research with human subjects. All experiments are conducted on existing public benchmark datasets.

\item {\bf Institutional review board (IRB) approvals or equivalent for research with human subjects}
    \item[] Question: Does the paper describe potential risks incurred by study participants, whether such risks were disclosed to the subjects, and whether Institutional Review Board (IRB) approvals (or an equivalent approval/review based on the requirements of your country or institution) were obtained?
    \item[] Answer: \answerNA{} %
    \item[] Justification: The paper does not involve crowdsourcing, user studies, or any direct research with human subjects, and therefore IRB approval or an equivalent review process was not required.

\item {\bf Declaration of LLM usage}
    \item[] Question: Does the paper describe the usage of LLMs if it is an important, original, or non-standard component of the core methods in this research? Note that if the LLM is used only for writing, editing, or formatting purposes and does \emph{not} impact the core methodology, scientific rigor, or originality of the research, declaration is not required.
    \item[] Answer: \answerYes{} %
    \item[] Justification: The paper uses an LLM-based evidence-bound extraction step as a non-standard preprocessing component of the method. Its role, scope, and prompt design are described in the methodology and appendix, where we clarify that the LLM is used to extract and normalize structured functional information rather than to generate the final recommendation outputs.

\end{enumerate}

\end{document}